\documentclass[aip,twocolumn,reprint,superscriptaddress]{revtex4-1}
\usepackage{amsmath}
\usepackage{amssymb}
\usepackage[latin9]{inputenc}
\usepackage{graphicx}
\usepackage{hyperref}
\usepackage{siunitx}
\usepackage{graphicx}
\usepackage[color= blue!20]{todonotes}
\usepackage[normalem]{ulem}
\usepackage{color}
\usepackage{ulem}
\usepackage[symbol]{footmisc}

\begin{document}

\onecolumngrid

\noindent\textbf{\textsf{\Large Cavity electromechanics with parametric mechanical driving}}

\normalsize
\vspace{.3cm}

\noindent\textsf{D.~Bothner$^{1, \dagger}$, S.~Yanai$^1$, A.~Iniguez-Rabago$^1$, M.~Yuan$^{1, 2}$, Ya.~M.~Blanter$^1$, and G.~A.~Steele$^1$}

\vspace{.2cm}
\noindent\textit{$^1$Kavli Institute of Nanoscience, Delft University of Technology, PO Box 5046, 2600 GA Delft, The Netherlands\\
$^2$Present address: Paul-Drude Institut f\"ur Festk\"orperelektronik Leibniz-Institut im Forschungsverband Berlin e.V., Hausvogteiplatz 5-7, 10117 Berlin, Germany\vspace{0.2cm}\\$^\dagger$\normalfont{d.bothner-1@tudelft.nl}}

\vspace{.5cm}

\date{\today}

{\addtolength{\leftskip}{10 mm}
\addtolength{\rightskip}{10 mm}

Microwave optomechanical circuits have been demonstrated in the past years to be extremely powerfool tools for both, exploring fundamental physics of macroscopic mechanical oscillators as well as being promising candidates for novel on-chip quantum limited microwave devices.
In most experiments so far, the mechanical oscillator is either used as a passive device element and its displacement is detected using the superconducting cavity or manipulated by intracavity fields.
Here, we explore the possibility to directly and parametrically manipulate the mechanical nanobeam resonator of a cavity electromechanical system, which provides additional functionality to the toolbox of microwave optomechanical devices.
In addition to using the cavity as an interferometer to detect parametrically modulated mechanical displacement and squeezed thermomechanical motion, we demonstrate that parametric modulation of the nanobeam resonance frequency can realize a phase-sensitive parametric amplifier for intracavity microwave photons.
In contrast to many other microwave amplification schemes using electromechanical circuits, the presented technique allows for simultaneous cooling of the mechanical element, which potentially enables this type of optomechanical microwave amplifier to be quantum-limited.

}

\vspace{.5cm}

\twocolumngrid

\section*{Introduction}
\vspace{-2mm}
%
%
Superconducting microwave circuits have been demonstrated to be extremely powerful tools for the fields of quantum information processing \cite{Wallraff04, DiCarlo09, Barends16}, circuit quantum electrodynamics \cite{Schuster07, Hofheinz08, Bosman17, Langford17, Gely19}, astrophysical detector technologies \cite{Day03} and microwave optomechanics \cite{Regal08, Teufel11, Aspelmeyer14}.
%
%
In the latter, microwave fields in superconducting cavities are parametrically coupled to mechanical elements such as suspended capacitor drumheads or metallized nanobeams, enabling high-precision detection and manipulation of mechanical motion.
%
%
Milestones achieved in the field include sideband-cooling of mechanical oscillators to the quantum ground state \cite{Teufel11}, strong coupling between photons and phonons \cite{Teufel11a}, the generation of non-Gaussian states of motion \cite{Wollman15, Pirkkalainen15, Reed17} or the entanglement between two mechanial oscillators \cite{OckeloenKorppi18}.

%
%
Recently, there are increasing efforts taken towards building passive and active quantum limited microwave elements for quantum technologies based on microwave optomechanical circuits, connecting the fields of microwave optomechanics, circuit quantum electrodynamics and quantum information science \cite{Metelmann14, Nunnenkamp14, Toth17}. 
%
%
Among the most important developments into this direction are the demonstration of microwave amplification by blue sideband driving in simple optomechanical circuits \cite{Massel11}, and the realization of directional microwave amplifiers \cite{Malz18} as well as microwave circulators \cite{Bernier17, Barzanjeh17} in more complex multimode systems \cite{OckeloenKorppi16}.

%
%
%
%
%
%
Recent theoretical work \cite{Levitan16, Lemonde16, Qvarfort19} on optomechanical systems with a parametrically driven mechanical oscillator proposed the use of mechanical parametric driving to enable parametric amplification with enhanced bandwidth and reduced added noise, compared to the case of a optomechanical amplifier using a blue-sideband drive \cite{Levitan16}. 
Furthermore, the authors predict that there is a parameter regime that results in an effective density of states, which can be interpreted as an effective negative temperature for cavity photons \cite{Levitan16}.
Other recent works have predicted enhancements of the optomechanical coupling \cite{Lemonde16} and the generation of non-Gaussian microwave states \cite{Qvarfort19}.
Direct electrostatic driving of a mechanical element in an microwave electromechnical cavity using a combination of DC fields and electrical fields resonant with the lower frequency mechanical device have been used in the past for probing mechanical resonators in cavity devices \cite{Regal08, Weber14, Singh14}.
These schemes also allow tuning of the mechanical frequency in an optomechanical cavity \cite{Weber14, Singh14, Andrews15} and enabling direct parametric driving of the mechanical resonator. Using this electrostatic tuning for parametric driving in optomechanics, however, has until now not been explored.
\begin{figure*}
\centerline{\includegraphics[trim = {1.5cm, 5.5cm, 1.5cm, 1cm}, clip=True, width=0.97\textwidth]{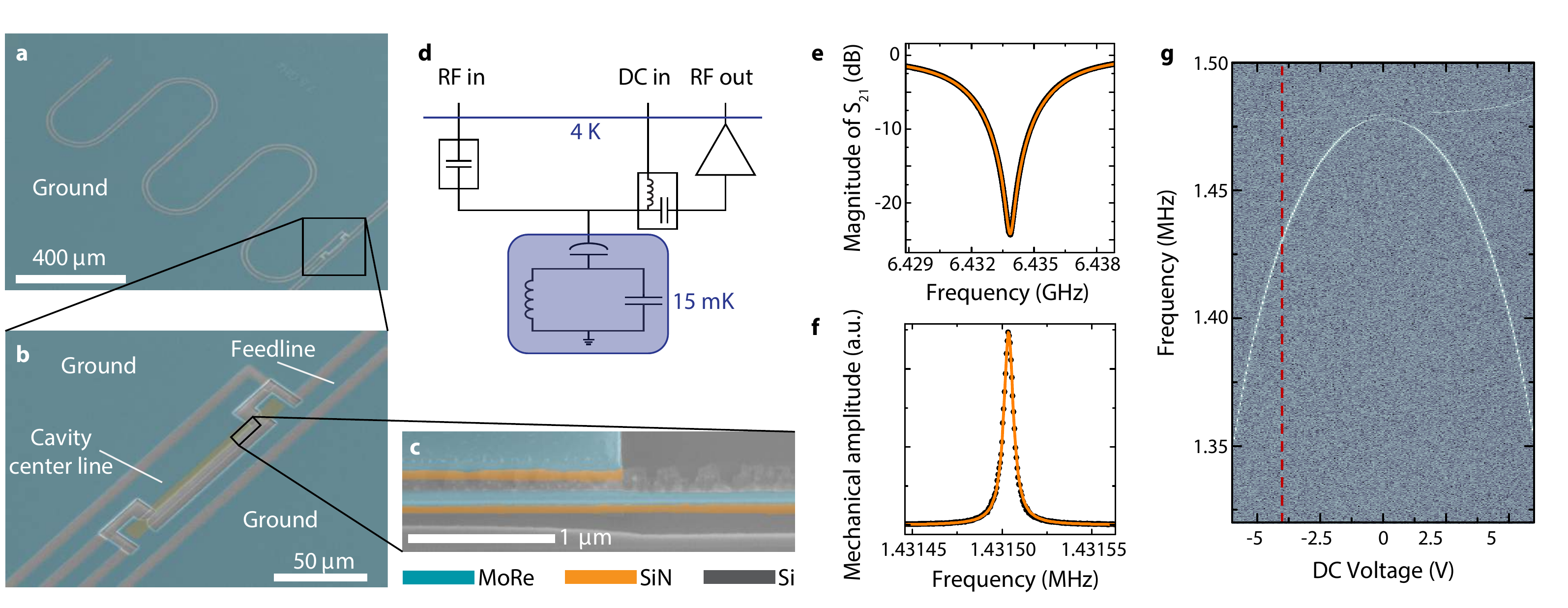}}
\caption{\textsf{\textbf{Superconducting circuit nano-electromechanical system with electrostatic and low-frequency access to the mechanical nanobeam.} \textbf{a}, False-color scanning electron microscopy image of a superconducting quarter-wavelength cavity (here for $\omega_c = 2\pi\cdot7.5\,$GHz), capacitively side-coupled to a coplanar waveguide feedline. The molybdenum-rhenium (MoRe) metallization is shown in blue and the silicon (Si) substrate in gray. \textbf{b}, Zoom into the coupling capacitance region, where the mechanical nanobeam as part of the coupling capacitance is visible. The dimensions of the beam, which consists of MoRe on top of high-stress silicon-nitride (Si$_3$N$_4$), are $100\,\mu$m$\,\times\,150\,$nm$\,\times\,143\,$nm. \textbf{c}, A magnified view of the suspended nanobeam. \textbf{d}, Simplified circuit and measurement scheme, showing a lumped element circuit representation of the device as well as the microwave input and output lines (including a DC block and high electron mobility transistor amplifier shown as triangle) and the DC input line connected to the microwave lines via a bias-tee. A more detailed version of the set-up is given in the Supplementary Information Sec.~S2. \textbf{e}, Cavity resonance data (black) and fit curve (orange). From the fit, we extract the cavity resonance frequency $\omega_c = 2\pi\cdot 6.434\,$GHz and the internal and external linewidths $\kappa_i = 2\pi \cdot 370\,$kHz and $\kappa_e = 2\pi\cdot 5.7\,$MHz, respectively. \textbf{f}, Resonance curve of the mechanical oscillator read-out via the superconducting cavity. Data are shown as black dots, a Lorentzian fit as orange line. From the fit we extract the mechanical resonance frequency $\Omega_m = 2\pi\cdot 1.4315\,$MHz and a quality factor $Q_m = 195000$. \textbf{g}, Optomechanically detected excitation spectrum of the nanobeam vs applied DC voltage. The bright line resembling an inverted parabola represents the resonance of the in-plane mode, which was used everywhere throughout this paper. The thin second line around $1.48\,$MHz corresponds to the mechanical out-of-plane mode. The red dashed line at $V_\mathrm{dc} = -4\,$V indicates the voltage operation point we chose to use.}}
\label{fig:device}
\end{figure*}
Here, we present measurements of a superconducting microwave optomechanical device in which we use direct electrostatic driving to achieve strong parametric modulation of the mechanical element.
By modulating the mechanical resonance frequency, we generate phase-sensitive parametric amplitude amplification and thermomechanical noise squeezing of the mechanical motion, both detected using optomechanical cavity interferometry \cite{Regal08}.
Furthermore, we demonstrate how parametric modulation of the mechanical resonance frequency can be used to generate phase-sensitive amplification of a microwave probe tone, which is three orders of magnitude larger in frequency than the parametric pump tone itself.
For the operation of the microwave amplifier, the optomechanical system can be driven on the red cavity sideband, which allows for simultaneous mechanical cooling and microwave amplification. 
The experimental implementation presented here provides a optomechanical platform for further exploration of phase-senstive quantum limited amplification and photon bath engineering using mechanical parametric driving.

\section*{Results}
\vspace{-2mm}

\subsection*{The device}
\vspace{-2mm}

Fig.~\ref{fig:device} show an image of a superconducting coplanar waveguide (CPW) quarter-wavelength ($\lambda/4$) resonator used as a microwave cavity.
The cavity is patterned from a $\sim 60\,$nm thick film of $60/40$ molybdenum-rhenium alloy (MoRe, superconducting transition temperature $T_c \sim 9\,$K \cite{Singh14a}) on a $10\times 10\,$mm$^2$ and $500\,\mu$m thick high-resistivity silicon substrate, cf. SM Sec.~S1.
For driving and readout, the cavity is capacitively side-coupled to a transmission feedline by means of a coupling capacitance $C_c = 16\,$fF.
The cavity has a fundamental mode resonance frequency $\omega_c = 2\pi\cdot 6.434\,$GHz and internal and external linewidths $\kappa_i = 2\pi\cdot 370\,$kHz and $\kappa_e = 2\pi\cdot 5.7\,$MHz, respectively.
The transmission spectrum of the cavity around its resonance frequency is shown in Fig.~\ref{fig:device}\textbf{e}, for details on the device modeling and fitting see Supplemetary Material Sec.~S3.
The superconducting cavity is parametrically coupled to a MoRe-coated high-stress Si$_3$N$_4$ nanobeam, which is electrically integrated into the transmission feedline. 
The nanobeam has a width $w = 150\,$nm, a total thickness $t = 143\,$nm (of which $\sim 83\,$nm are Si$_3$N$_4$ and $60\,$nm are MoRe) and a length $r = 100\,\mu$m.
It is separated from the center conductor of the cavity by a $\sim 200\,$nm wide gap, cf. Fig.~\ref{fig:device}\textbf{c}.
More design and fabrication details are described in the Supplementary Material Sec.~S1.
The mechanical nanobeam oscillator has a resonance frequency of its fundamental in-plane mode of $\Omega_{m0} = 2\pi\cdot 1.475\,$MHz.
It can be significantly tuned by applying a DC voltage $V_\mathrm{dc}$ between center conductor and ground of the coplanar waveguide feedline, adding an electrostatic spring constant to the intrinsic spring, cf. Supplementary Material Sec.~S5.
The measured functional dependence of the resonance frequency on DC voltage is shown in Fig.~\ref{fig:device}\textbf{g}.
Throughout this whole article, we bias the mechanical resonator with $V_\mathrm{dc} = -4\,$V, leading to a resonance frequency $\Omega_m = 2\pi\cdot 1.4315\,$MHz and a linewidth $\Gamma_m \approx 2\pi\cdot7.5\,$Hz.
A resonance curve of the mechanical oscillator at $V_\mathrm{dc} = -4\,$V is shown in Fig.~\ref{fig:device}\textbf{f}.
The device is operated in a dilution refrigerator with a base temperature of $T_b = 15\,$mK, which corresponds to a thermal cavity occupation of $\frac{k_\mathrm{B}T_b}{\hbar\omega_c} \sim 0.05$ photons.
Assuming the mode temperature of the nanobeam being the fridge base temperature, we expect an average occupation of the mechanical mode with $n_m = k_\mathrm{B}T_m/\hbar\Omega_m \sim 220$ thermal phonons.

\subsection*{Parametric mechanical amplitude amplification}
\vspace{-2mm}

\begin{figure}
\centerline{\includegraphics[trim = {0.5cm, 6.5cm, 0.5cm, 0.5cm}, clip=True, width=.48\textwidth]{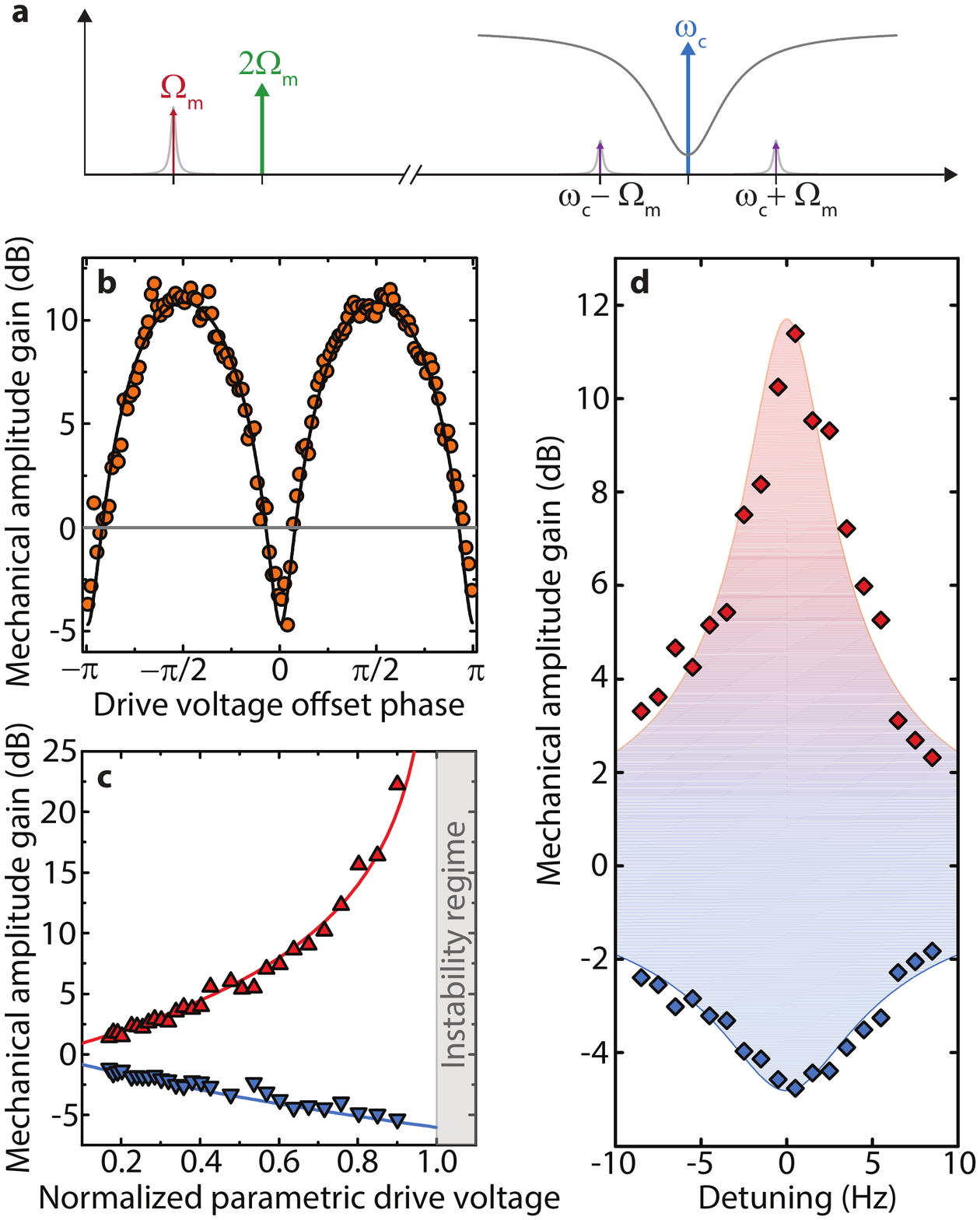}}
\caption{\textsf{\textbf{Optomechanical detection of parametric, phase-sensitive mechanical amplitude amplification by means of modulating the electrostatic spring constant.} \textbf{a}, Experimental scheme. The mechanical oscillator is coherently driven by a combination of DC and alternating voltage with frequency $\Omega \sim \Omega_m$, while the electrostatic spring constant is modulated with twice this frequency $2\Omega\sim 2\Omega_m$. Via the optomechanical coupling, the mechanical oscillations generate sidebands to a microwave pump tone sent to the cavity with frequency $\omega = \omega_c$, which are used for homodyne detection of the mechanical amplitude. \textbf{b}, Mechanical amplitude gain vs offset phase $\phi_p$ between resonant drive and parametric modulation. When the phase is swept, the amplitude is oscillating between amplification or de-amplification with a periodicity of $\pi$. Circles show data and the line shows a fit with the theoretical expression Eq.~(\ref{eqn:gain}). \textbf{c}, Maximum and minimum gain on resonance vs parametric modulation strength. The maximum ($\phi_p = \pi/2$) and minimum ($\phi_p = 0$) gain values on resonance follow the theoretical curves (lines) up to a maximum gain of $\sim22\,$dB. For stronger parametric modulation amplitudes close to the instability threshold (indicated as vertical line), the gain in our experiments is limited by resonance frequency fluctuations of the mechanical resonator. \textbf{d}, Maximum and minimum gain vs detuning from resonance. For a driving frequency slightly detuned from resonance, the maximum gain gets reduced compared to the resonant case. Points are extracted from phase-sweep curve fits. Lines show the corresponding theoretical curves and the shaded area contains all gain values achievable by changing $\phi_p$.}}
\label{fig:amplification}
\end{figure}

%
When the resonance frequency $\Omega_m$ of a harmonic oscillator is modulated with twice the resonance frequency $\Omega_p = 2\Omega_m$, then a small starting amplitude of the oscillator motion can be increased or reduced, depending on the relative phase between the oscillator motion and the frequency modulation \cite{Rugar91}. 
To modulate the resonance frequency of a mechanical oscillator, one of the relevant system parameters like the oscillator mass $m$ or the restoring spring force constant $k$ can be modulated.
Here, we follow the latter approach and modulate the effective spring constant of the nanobeam by applying a combination of a static voltage $V_\mathrm{dc}$ and an oscillating voltage $V_{2\Omega}\cdot\sin{2\Omega t}$ with roughly twice the mechanical resonance frequency $\Omega \sim \Omega_m$.
The static voltage adds an electrostatic spring contribution $k_\mathrm{dc}$ to the intrinsic spring constant $k_m$ and the oscillating part modulates the total spring constant with $\sim 2\Omega_m$.
In addition, we slightly excite the mechanical oscillator by adding a near-resonant oscillating voltage $V_0\cos{(\Omega t + \phi_p)}$ and characterize its steady-state displacement amplitude depending on the parametric modulation amplitude $V_{2\Omega}$ and on the relative phase difference between resonant drive and parametric modulation $\phi_p$.
The mechanical amplitude is detected by monitoring the optomechanically generated sidebands to a microwave drive tone sent into the cavity, which is constant in amplitude and frequency with $\omega \sim \omega_c$, cf. Fig.~\ref{fig:amplification}\textbf{a}.
We operate the nanobeam in a regime of voltages where it can be modelled by the equation of motion
\begin{equation}
\ddot{x} + \Gamma_m \dot{x} +\frac{1}{m}\left[k_0 + k_p\sin{2\Omega t}\right]x = \frac{F_0}{m}\cos{\left(\Omega t + \phi_p\right)}
\end{equation}
where $m$ is the effective nanowire mass, $x$ is the effective nanowire displacement, $k_0 = k_m + k_\mathrm{dc}$, $k_p \propto V_\mathrm{dc}V_{2\Omega}$ and $F_0 \propto V_\mathrm{dc}V_0$.
From an approximate solution of this equation of motion, the parametric amplitude gain $G_p = |x|_\mathrm{on}/|x|_\mathrm{off}$ can be derived to be given by
\begin{equation}
G_p = \left[\frac{\cos^2{(\phi_p + \varphi)}}{\left(1 + \frac{V_{2\Omega}}{V_t}\right)^2} + \frac{\sin^2{(\phi_p + \varphi)}}{\left(1 - \frac{V_{2\Omega}}{V_t}\right)^2}\right]^{1/2}.
\label{eqn:gain}
\end{equation}
The detuning dependent threshold voltage $V_t$ for parametric instability in this relation is given by
\begin{equation}
V_t = V_{t0}\sqrt{1 + \frac{4\Delta_m^2}{\Gamma_m^2}}
\end{equation}
with the threshold voltage on resonance $V_{t0}$ and the detuning from mechanical resonance $\Delta_m = \Omega - \Omega_m$.
The phase $\varphi = -\arctan(2\Delta_m/\Gamma_m)$ considers the detuning dependent phase difference between the near-resonant driving force and the mechanical motion.
Details on the theoretical treatment of the device are given in the Supplementary Material Sec.~S7.
Figure~\ref{fig:amplification} summarizes our results on the phase and detuning dependent parametric frequency modulation.
When we excite the mechanical resonator exactly on resonance, apply a parametric modulation with twice the resonance frequency and sweep the phase $\phi_p$, we find an oscillatory behaviour between amplitude amplification and de-amplification with a periodicity of $\Delta\phi_p = \pi$, cf. Fig.~\ref{fig:amplification}\textbf{b}.
To explore the dependence of the amplification on the parametric modulation amplitude $V_{2\Omega}$, we repeat this experiment for different voltages $V_{2\Omega}$ and extract maximum and minimum gain by fitting the data with Eq.~(\ref{eqn:gain}) for $V_t = V_{t0}$ and $\varphi = 0$.
The extracted values follow closely the theoretical curves up to a voltage $V_{2\Omega} \approx 0.9 V_{t0}$, above which we are limited by resonance frequency fluctuations of the mechanical resonator.
The maximum gain we achieve by this is about $\sim 22\,$dB.
In order to characterize the device response also for drive frequencies detuned from resonance, we repeat the above measurements for different detunings and extract the maximum and minimum gain for each of these data sets.
Hereby, we always keep the parametric drive frequency twice the excitation frequency and not twice the resonance frequency. 
The maximum and minimum values of gain we find for $V_{2\Omega} \approx 0.75 V_{t0}$ are shown in Fig.~\ref{fig:amplification}~\textbf{d} and are in good agreement with theoretical curves shown as lines.
We note, that the dependence of maximum and minimum gain of detuning is not Lorentzian lineshaped, as the threshold voltage is detuning dependent itself and the deviations between experimental data and theoretical lines mainly occur due to slow and small resonance frequency drifts of the nanobeam.
Moreover, the phase between near-resonant excitation drive and parametric modulation for maximum/minimum gain does not have a constant value, it follows an $\arctan$-function as is discussed in more detail in the Supplementary Material.
In summary, we have achieved an excellent experimental control and theoretical modelling regarding the parametric amplification of the coherently driven nanobeam in both parameters, the relative phase between the drives and the detuning from mechanical resonance.

\subsection*{Thermomechanical noise squeezing}

\begin{figure}
\centerline{\includegraphics[trim = {1cm, 1cm, 1cm, 0cm}, clip=True, width=.48\textwidth]{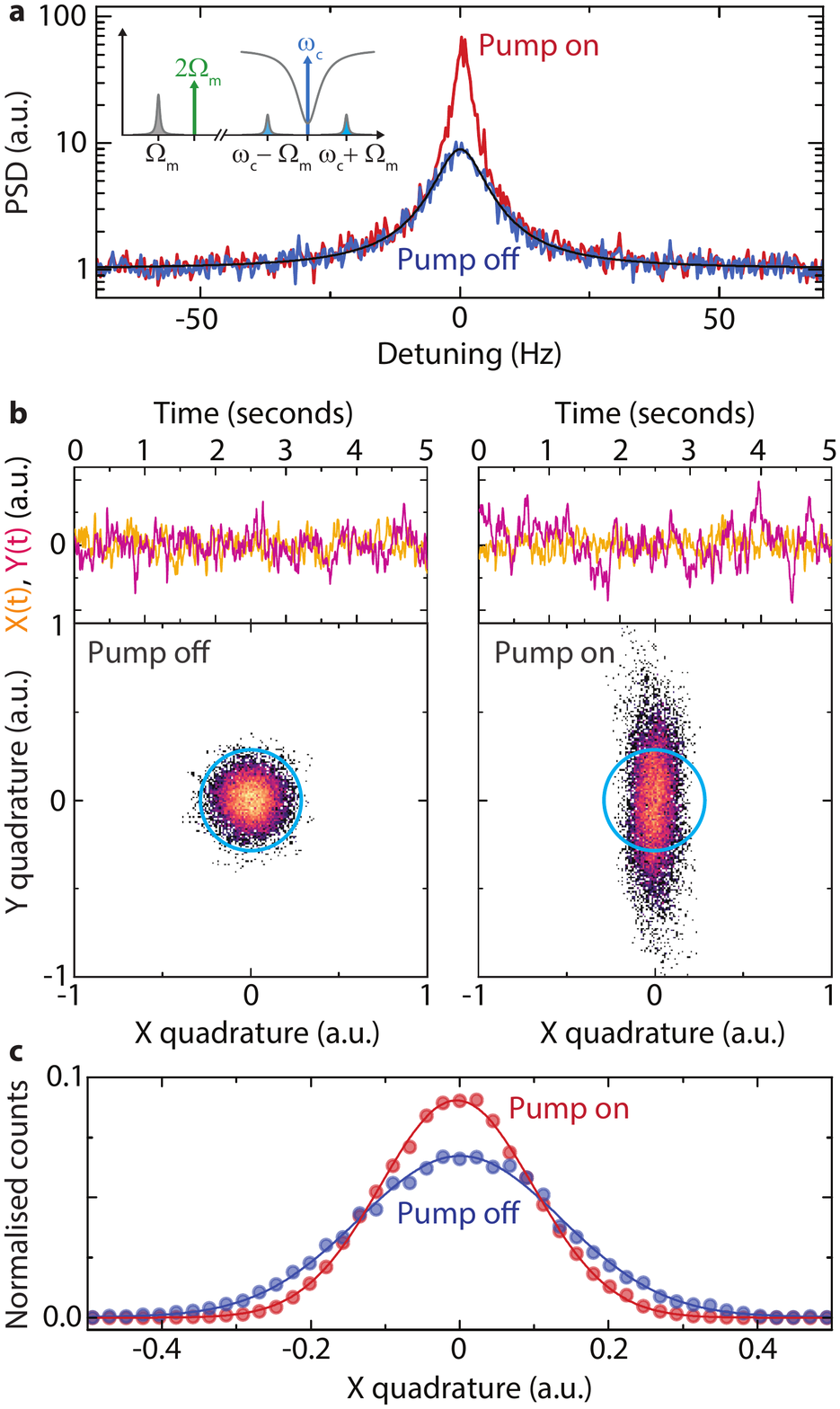}}
\caption{\textsf{\textbf{Interferometric detection of squeezed thermomechanical noise in a nanomechanical oscillator.} \textbf{a}, The thermal displacement fluctuations generate sidebands at $\omega = \omega_c + \Omega_m$ and $\omega = \omega_c - \Omega_m$ to a microwave tone send to the cavity at $\omega = \omega_c$ as schematically shown in the inset. After down-conversion, we detect these sidebands and the corresponding power spectral density is shown for the parametric modulation switched off as blue line and with the parametric modulation switched on as red line. The black line is a Lorentzian fit to the data without parametric modulation. \textbf{b} shows the quadratures of the thermal displacement fluctuations vs time in the top panels and as histograms (taken for $300\,$s of measurement time) in the bottom panels. Without parametric modulation, the thermal fluctuations are distributed equally in both quadratures (left side) and the quadrature histogram is a rotational symmetric Gaussian curve; with a parametric modulation applied, as shown on the right side, the fluctuations in one quadrature get amplified while the fluctuations in the second quadrature get de-amplified. The result is a squeezed thermal state. The colourscale represents histogram counts from low (dark) to high (orange) values. White pixels correspond to no recorded counts. The blue circles in the histogram plots are guides to the eye. In \textbf{c} we plot the distribution of $X$-quadrature values for the histograms shown in \textbf{b} as dots and Gaussian fits as lines. When the parametric modulation is switched on, the variance of the $X$-quadrature gets significantly decreased and the squeezing factor is approximately $s = 0.49$. The histograms are normalized to the total number of $\sim13000$ data points}}
\label{fig:squeezing}
\end{figure}
\begin{figure*}
\centerline{\includegraphics[trim = {0.0cm, 4.7cm, 3.6cm, 0.0cm}, clip=True, width=0.95\textwidth]{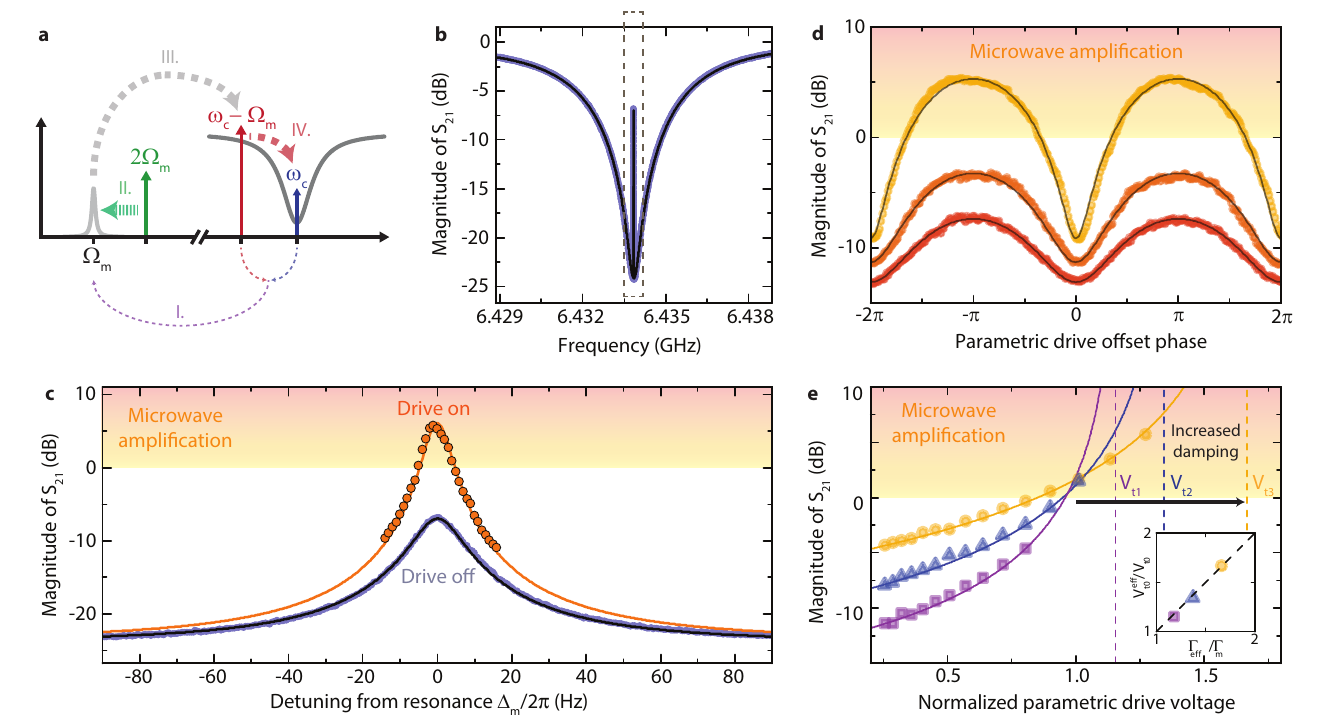}}
\caption{\textsf{\textbf{Phase-sensitive and tunable amplification of microwaves by parametrically pumping a nanomechanical oscillator in an optomechanical system.} \textbf{a}, Experimental scheme. The cavity is coherently driven by a strong drive tone on the red sideband $\omega_d = \omega_c - \Omega_m$. In addition a small probe tone is swept through the cavity resonance with $\omega_p = \omega_d + \Omega \sim \omega_c$ where $\Omega \sim \Omega_m$. At the same time, the resonance frequency of the mechanical oscillator is parametrically modulated with $2\Omega$. \textbf{b}, Optomechanically induced transparency (OMIT) without parametric modulation $V_{2\Omega} = 0$. By the presence of the red-sideband drive tone, an amplitude beating generated by interference with the probe tone drives the mechanical nanobeam around its resonance frequency, which in turn modulates the drive tone and generates sidebands. The upper of these sidebands interferes with the probe tone, opening a narrow transparency window in the absorption dip of the cavity. Data is shown in blue and black line is a fit, and the dashed box indicates the zoom-in region shown in \textbf{c}. \textbf{c}, Zoom into the OMIT transparency window. In addition to the data without parametric modulation, we show the highest achieved transmission with the parametric drive switched on as orange circles. Close to the mechanical resonance we observe gain of the microwave input probe signal up to $\sim 7\,$dB. The orange line shows a theoretical curve calculated with independently obtained system parameters. The schematic shown in \textbf{d} visualizes the amplification mechanism. By the beating of the two cavity tones, energy from the cavity field is converted into mechanical motion. This mechanical motion is amplified by means of parametric modulation and the hereby increased energy is upconverted back to the probe tone frequency by sideband generation of the red-sideband tone. \textbf{d} The microwave gain is phase-sensitive, i.e., it depends on the relative phase between the parametric modulation and the amplitude beating in the cavity. The three datasets (black lines are fits) show the gain for different detunings from $\omega_p - \omega_d = \Omega_m$ ($0\,$Hz, $7\,$Hz and $12\,$Hz). \textbf{e} Probe tone gain vs parametric drive voltage for three different red sideband drive powers. The parametric drive voltage is normalized to its value obtained in Fig.~2 using a resonant drive for amplitude detection. Lines are theoretical calculations based on independently extracted system parameters. The parametric instability threshold, indicated by dashed vertical lines, is shifted to higher values with increasing red-sideband drive power, partly due to optical damping, partly due to a power-dependent intrinsic mechanical damping rate. The inset shows the extracted threshold voltage vs effective mechanical linewidth and as dashed line the theoretical prediction.}}
\label{fig:OMIT}
\end{figure*}
Due to a large residual occupation of the mechanical mode with $10^2-10^3$ thermal phonons, its displacement is subject to thermal fluctuations, which in a narrow bandwidth can be described by \cite{Rugar91}
\begin{equation}
x_\mathrm{th}(t) = X(t)\cos{\Omega_m t} + Y(t)\sin{\Omega_m t}.
\end{equation}
Here, $X(t)$ and $Y(t)$ are random variable quadrature amplitudes, which vary slowly compared to $\Omega_m^{-1}$.
Similarly to the coherently driven mechanical amplitude detection discussed above, this thermal motion or thermomechanical noise can be measured by optomechanical sideband generation in the output field of a microwave signal sent into the superconducting cavity, cf. the inset schematic in Fig.~\ref{fig:squeezing}\textbf{a}.
We measure the thermomechanical noise quadratures $X(t)$ and $Y(t)$ with and without parametric pump.
An exemplary result is shown in Fig.~\ref{fig:squeezing}\textbf{b}.
As we have demonstrated above by amplification and de-amplification of a coherent excitation, one of the quadrature amplitudes, here $Y(t)$, is getting amplified while the other, here $X(t)$, is simultaneously reduced, when the mechanical resonance frequency is parametrically modulated with $2\Omega_m$.
This puts the mechanical nanobeam into a squeezed thermal state.
From the time traces of the quadratures, we reconstruct by means of a Fourier transform the power spectral density of the noise as shown in Fig.~\ref{fig:squeezing}\textbf{a}.
With parametric driving, the total power spectral density is larger than without, in particular close to $\Omega_m$, as the additional energy pumped into the amplified quadrature $Y(t)$ is larger than the energy reduction in $X(t)$ and at the same time the total linewidth decreases for the same reason.
From the time traces, we can also generate quadrature amplitude histograms, shown in the bottom panels of Fig.~\ref{fig:squeezing}\textbf{b}.
In the histograms the squeezing of the thermal noise is apparent as a deformation from a circular, two-dimensional (2D) Gaussian distribution in the case without parametric pump to a cigar-like shaped overall probability distribution, when the parametric modulation is applied.
To determine the squeezing factor we achieve by this, we integrate the 2D-histrograms along the $Y$-quadrature and extract the variance $\sigma_{x}^2$ of the $X$-quadrature from a Gaussian fit to the resulting data, cf. Fig.~\ref{fig:squeezing}\textbf{c}.
For the parametric modulation amplitudes used here, we find the squeezing factor
\begin{equation}
s = \frac{\sigma_{x,\mathrm{on}}^2-\sigma_\mathrm{amp}^2}{\sigma_{x, \mathrm{off}}^2 - \sigma_\mathrm{amp}^2} = 0.49,
\end{equation}
where $\sigma_{x, \mathrm{on}}^2$ and $\sigma_{x, \mathrm{off}}^2$ are the $X$-quadrature variances with the parametric drive on and off, respectively.
The variance $\sigma_\mathrm{amp}^2$ is the quadrature noise originating from the cryogenic amplifier in our detection chain and is measured by monitoring the noise slightly detuned from the mechanical resonanace.
More details about the measurement scheme and data processing can be found in the Supplementary Material Sec.~S9.
\subsection*{Parametric microwave amplification}
In a cavity optomechanical system, the mechanical oscillator can not only be coherently driven by a directly applied resonant force, but also by amplitude modulations of the intracavity field.
Such a near-resonant amplitude modulation can be generated by sending two microwave tones with a frequency difference close to the mechanical resonance into the cavity.
Here, we apply a strong microwave drive tone on the red sideband of the cavity, i.e., at $\omega_d = \omega_c - \Omega_m$, and add a small probe signal around the cavity resonance frequency at $\omega_p \sim \omega_c$.
This experimental scheme generates a phenomenon called optomechanically induced transparency (OMIT), where by interference a narrow transparency window opens up in the center of the cavity absorption dip \cite{Agarwal10, Weis10}.
The width of the transparency window is given by the sum of intrinsic mechanical linewidth $\Gamma_m$ and the additional linewidth due to the red sideband drive-induced optical damping $\Gamma_o$.
The effect of OMIT effect can be understood as follows.
The amplitude beating between the two microwave tones coherently drives the nanobeam by an oscillating radiation pressure force, which transfers energy from the cavity field to the nanobeam.
The resulting mechanical motion with frequency $\Omega = \omega_p - \omega_d$ modulates the cavity resonance frequency and hereby generates sidebands to the intracavity drive tone at $\omega_d \pm \Omega$, with a well-defined phase relation to the probe tone.
The sideband generated at $\omega_d + \Omega$ interferes with the probe signal and generates OMIT, cf. Fig.~\ref{fig:OMIT}\textbf{a} (for vanishing parametric modulation) and \textbf{b}.
In \textbf{b}, the transparency window can be seen in the center of the cavity transmission spectrum as extremely narrow spectral line and a zoom into this region, shown in \textbf{c}, reveals the Lorentzian lineshape with a width $\Gamma_\mathrm{eff} \approx 2\pi\cdot 12\,$Hz.
When we perform the OMIT protocol with a parametric modulation applied to the nanobeam, the mechanical oscillations get modified according to the previously shown results, i.e., dependent on the relative phase between the cavity field-induced mechanical oscillation and the parametric modulation, the mechanical amplitude gets amplified or de-amplified.
By choosing the optimal phase for each detuning $\Delta_m = \Omega - \Omega_m$, the transparency window amplitude can be increased to values above $1$, i.e., the microwave probe tone is amplified by parametrically pumping the mechanical resonator, which is three orders of magnitude smaller in frequency than the probe signal, cf. Fig.~\ref{fig:OMIT}\textbf{c}.
With an amplified mechanical motion, the motion-induced sideband of the drive tone gets amplified as well, such that the total cavity output field at the probe frequency can be enhanced to values larger than $1$. 
A schematic of OMIT and the amplification mechanism is shown in Fig.~\ref{fig:OMIT}\textbf{a}.
The observed microwave amplification is, similar to the bare mechanical amplitude gain, phase-sensitive and modulates between amplification and de-amplification when sweeping the phase of the parametric drive, with a periodicity of $2\pi$.
This phase-sensitivity of the microwave gain is shown in Fig.~\ref{fig:OMIT}\textbf{d} for three different detunings from the mechanical resonance.
We note, that the phase periodicity here is equivalent to the case of the mechanical amplitude amplification, but due to the details of our theoretical analysis of the system (see SM Sec.~S10) the phase is given for the parametric drive instead of the resonant force here, which doubles its value.
Similar to the mechanical amplitude amplification, the microwave gain depends on the parametric drive voltage, which has a threshold value above which the parametric instability regime begins.
When we plot the maximally achievable transmission $|S_{21}|$ exactly on the mechanical resonance vs the parametric excitation voltage, we find a monotonously increasing behaviour as shown in Fig.~\ref{fig:OMIT}\textbf{e} for three different red-sideband drive powers (shown are powers corresponding to cooperativities $C_1\sim 0.16$, $C_2\sim 0.28$ and $C_3\sim 0.5$).
The functional dependence of the maximum transmitted power is formally identical to the case without parametric driving
\begin{equation}
|S_{21}|^2 = \frac{\kappa_i^2}{\kappa^2} + C_p\frac{\Gamma_m}{\Gamma_\mathrm{eff}^2}\left[2\frac{\kappa_i\kappa_e}{\kappa^2}\Gamma_\mathrm{eff} + \frac{\kappa_e^2}{\kappa^2}C_p\Gamma_m \right]
\end{equation}
with a parametrically enhanced cooperativity
\begin{equation}
C_p = \frac{C}{1-\frac{V_{2\Omega}}{V_{t0}^\mathrm{eff}}}
\end{equation}
where the effective threshold voltage is given by $V_{t0}^\mathrm{eff} = V_{t0}\Gamma_\mathrm{eff}/\Gamma_m$.
From fits to the data, shown as lines, we can extract the instability threshold voltages, indicated as dashed vertical lines and plotted in the inset vs effective mechanical linewidth.
The threshold gets shifted towards higher values due to an increase of mechanical linewidth, which is partly due to the optical spring and partly due to a microwave power-dependent intrinsic linewidth, see Supplementary Material Sec.~S6.
At the same time, the net microwave gain increases with increasing sideband drive power, as the baseline (the peak height of the transparency window) is shifted up as well and because the gain in this experiment was limited by the mechanical nonlinearity, which gets less significant for a larger total mechanical linewidth.

\section*{Conclusion}
\vspace{-2mm}

In this work, we have demonstrated an electromechanical cavity with mechanical parametric driving.  
By means of an optomechanical, interferometric readout scheme of a high quality factor mechanical nanobeam oscillator, we have demonstrated phase-sensitive mechanical amplitude amplification, and observed thermomechanical noise squeezing.
We demonstrated that this parametric mechanical drive can be used to implement a phase-sensitive microwave amplification, in a regime where dynamical backaction can simultaneously cool the mechanical resonator.
Using this new experimental platform in an optimized device, it should be possible to cool the mechanical oscillator into its quantum ground state and perform a near quantum-limited amplification scheme for microwave photons.
Furthermore will the presented approach allow to explore novel regimes of bath engineering of microwave cavities\cite{Levitan16} and enable other novel applications of mechnaical parametric driving and mechanical squeezing \cite{Lemonde16, Qvarfort19}.

\subsection*{Acknowledgements}
\vspace{-2mm}

This research was supported by the Netherlands Organisation for Scientific Research (NWO) in the Innovational Research Incentives Scheme -- VIDI, project 680-47-526.
This project has received funding from the European Research Council (ERC) under the European Union's Horizon 2020 research and innovation programme (grant agreement No 681476 - QOMD) and from the European Union's Horizon 2020 research and innovation programme under grant agreement No 732894 - HOT.

\subsection*{Author contributions}
\vspace{-2mm}

AIR, SY and GAS designed and fabricated the device.
DB, SY and MY performed the measurements.
DB, SY and GAS analysed the data.
DB and YMB developed the theoretical treatment.
GAS conceived the experiment and supervised the project.
DB wrote the manuscript with input from GAS.
All authors discussed the results and the manuscript.

\subsection*{Competing financial interest}
\vspace{-2mm}
The authors declare no competing financial interests.

\clearpage

\widetext

\noindent\textbf{\textsf{\Large Supplementary Information: Cavity electromechanics with parametric\\ mechanical driving}}

\normalsize
\vspace{.3cm}

\noindent\textsf{D.~Bothner$^{1, \dagger}$, S.~Yanai$^1$, A.~Iniguez-Rabago$^1$, M.~Yuan$^{1, 2}$, Ya.~M.~Blanter$^1$, and G.~A.~Steele$^1$}

\vspace{.2cm}
\noindent\textit{$^1$Kavli Institute of Nanoscience, Delft University of Technology, PO Box 5046, 2600 GA Delft, The Netherlands\\
$^2$Present address: Paul-Drude Institut f\"ur Festk\"orperelektronik Leibniz-Institut im Forschungsverband Berlin e.V., Hausvogteiplatz 5-7, 10117 Berlin, Germany\vspace{0.2cm}\\$^\dagger$\normalfont{d.bothner-1@tudelft.nl}}

\renewcommand{\thefigure}{S\arabic{figure}}
\renewcommand{\theequation}{S\arabic{equation}}

\renewcommand{\thesection}{S\arabic{section}}
\renewcommand{\bibnumfmt}[1]{[S#1]}

\setcounter{figure}{0}
\setcounter{equation}{0}

\section{Device fabrication}

\begin{figure}[h]
	\centering {\includegraphics[trim={0cm 9cm 0cm 1cm},clip=True,scale=0.87]{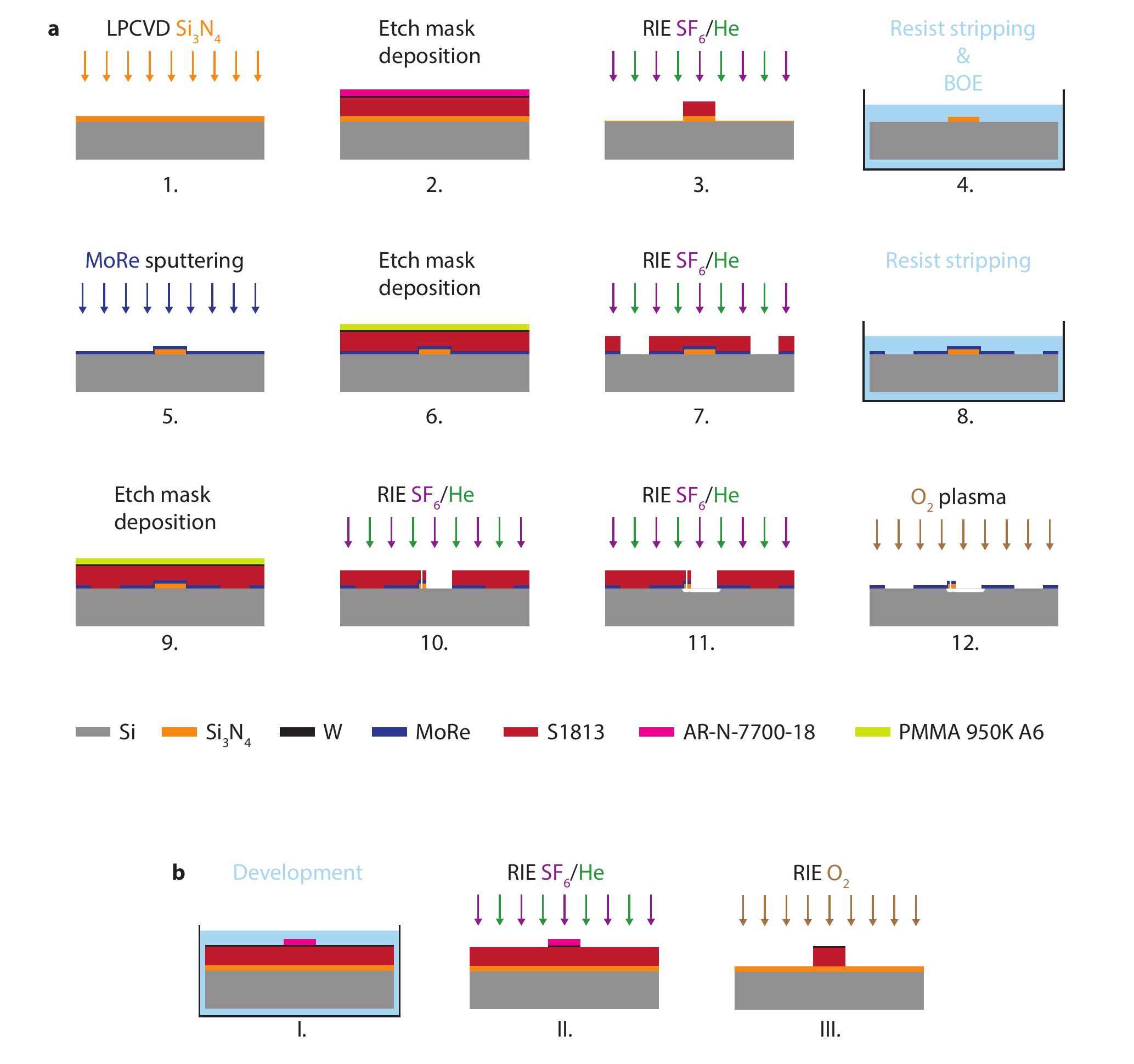}}
	\caption{\textsf{\textbf{Schematic device fabrication.} \textbf{a} 1.-4. show the deposition and patterning of the Si$_3$N$_4$ patches, 5.-8. show the deposition and patterning of the superconducting microwave structures and 9.-12. show the nanobeam patterning and release. \textbf{b} Steps between 2. and 3. of a. Equivalent steps are performed between 6. and 7. and between 9. and 10. of a. Dimensions are not to scale. A description of the individual steps is given in the text.}}
	\label{fig:Fab}
\end{figure}

The device fabrication started with the deposition of a $100\,$nm thick layer of high-stress Si$_3$N$_4$ on top of a $500\,\mu$m thick two inch silicon wafer by means of low pressure chemical vapour deposition (LPCVD). 
Afterwards, $60\,$nm thick gold markers on a $10\,$nm chromium adhesion layer were patterned onto the wafer using electron beam lithography (EBL), electron beam evaporation of the metals and lift-off.
Then, the wafer was diced into individual $10\times10\,$mm$^2$ chips, which were used for the subsequent fabrication steps.
By using a three-layer mask (S1813, tungsten and ARN-7700-18), EBL and several reactive ion etching (RIE) steps with O$_2$ and a SF$_6$/He gas mixture, the Si$_3$N$_4$ was thinned down everywhere to $\sim10\,$nm on the chip surface except for rectangular patches ($124\times9\,\mu$m large) around the future locations of the nanobeams.
After resist stripping in PRS3000, the remaining $\sim 10\,$nm of Si$_3$N$_4$ were removed in a buffered oxide etching (BOE) step, which also thinned down the Si$_3$N$_4$ in the rectangular patch areas to $\sim 83\,$nm.
This two-step removal of Si$_3$N$_4$ by dry and wet etching was performed in order to avoid over-etching with RIE into the silicon substrate.
Immediately afterwards, a $\sim60\,$nm thick layer of superconducting molybdenum-rhenium alloy (MoRe, 60/40) was sputtered onto the chip.
By means of another three-layer mask (S1813, W, PMMA 950K A6), EBL, O$_2$ and SF$_6$/He RIE, the microwave structures were patterned into the MoRe layer.
The remaining resist was stripped off in PRS3000.
Finally, the nanobeam patterning and release was performed.
The pattern definition was done using another three-layer mask (S1813, W, PMMA 950K A6), EBL and RIE.
After the MoRe-Si$_3$N$_4$ bilayer was completely etched by the SF$_6$/He gas mixture, the etching was continued for several minutes.
As we had chosen the RIE parameters to achieve slight lateral etching, the silicon underneath the narrow nanobeam was etched away by this measure and the beam was released from the substrate.
After the nanobeam release, the remaining resist was stripped using an O$_2$ plasma.
A simplified schematic of the fabrication is shown in Fig.~\ref{fig:Fab}, omitting the patterning of the electron beam markers.
\begin{figure}
	\centering {\includegraphics[trim={0cm 0.5cm 1.5cm 0.5cm},clip=True,scale=0.8]{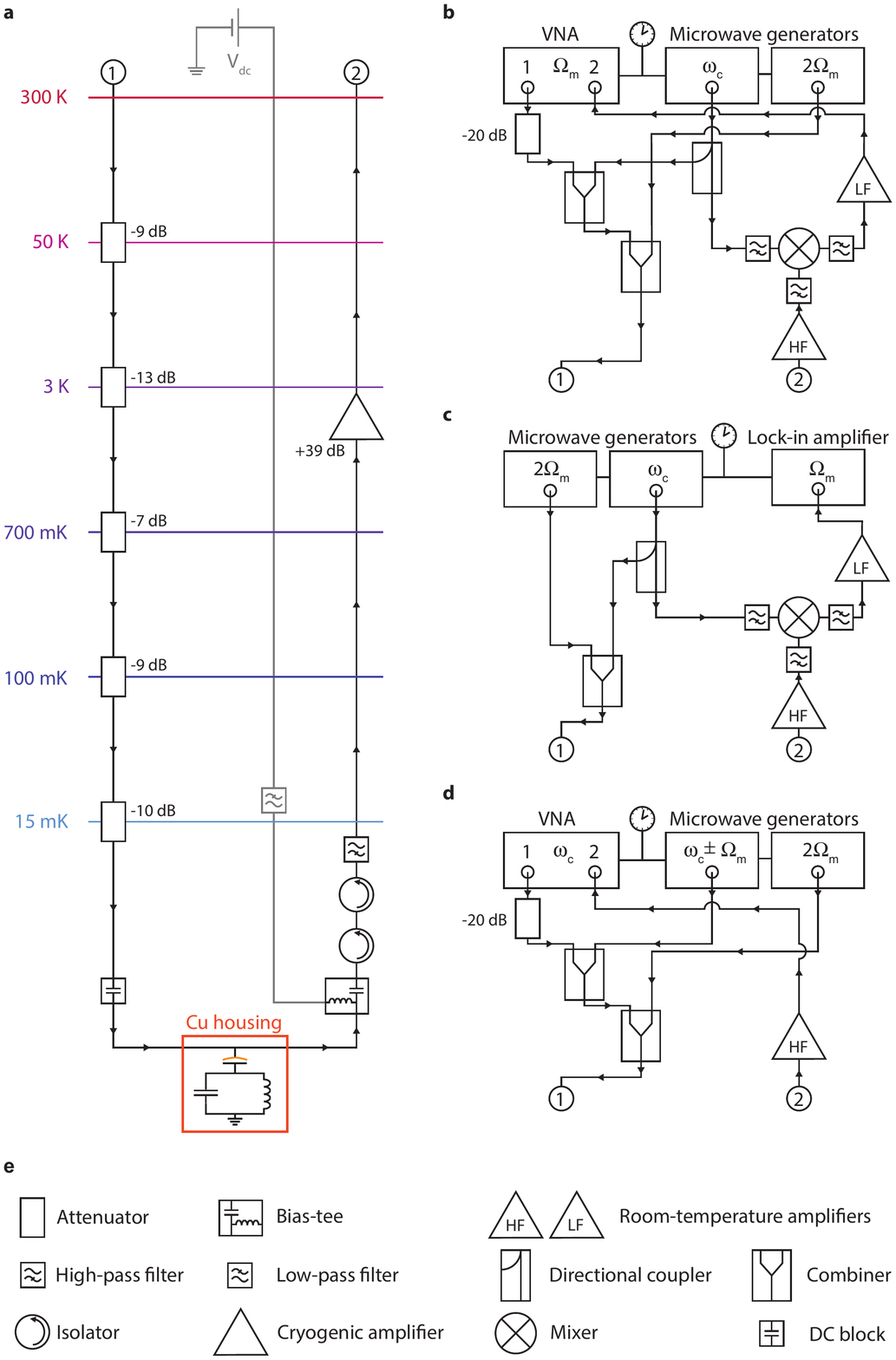}}
	\caption{\textsf{\textbf{Schematic of the measurement setup.} Details are given in the text.}}
	\label{fig:Setup}
\end{figure}

\section{Measurement setup}

Figure~\ref{fig:Setup} shows a schematic of the measurement setup configurations, which we used for the experiments reported in this paper.
All experiments were carried out in a dilution refrigerator with base temperature $T_b = 15\,$mK, cf. Fig.~\ref{fig:Setup}\textbf{a}.
The sample was mounted into a radiation tight copper housing and connected to two coaxial high-frequency lines.
By means of two bias-tees, the center conductors of the coaxial cables were also connected to DC wires and a DC voltage source, which allowed for DC access to the sample.
The input line was heavily attenuated to equilibrate the thermal radiation on the line to the refrigerator base temperature.
To isolate the sample from the noise of the cryogenic amplifier on the output line, we used two isolators in series on the milliKelvin plate.
Outside of the refrigerator, we used different configurations of microwave signal sources and high-frequency electronics for the three experiments presented here.
All three are shown in Figs.\ref{fig:Setup}\textbf{b},\textbf{c} ,and \textbf{d}, where the setup for the mechanical parametric amplification experiment is shown in \textbf{b}, the setup for the thermomechanical noise squeezing is shown in \textbf{c} and the configuration for the parametric microwave amplification in \textbf{d}.
For all experiments, the microwave sources and vector network analyzers (VNA) as well as the lock-in amplifier used a single reference clock of one of the devices.
Figure\ref{fig:Setup}\textbf{e} provides a symbol legend for \textbf{a} to \textbf{d}.

\section{Cavity characterization}

\subsection{The cavity model}

The cavity used in this experiment is a quarter-wavelength ($\lambda/4$) transmission line cavity, capacitively side-coupled to a microwave feedline via a coupling capacitor $C_c$ at the open end and shorted to ground at the other end.
Cavity and feedline have both the characteristic impedance $Z_0$ and the cavity has length $l$ and resonance frequency $\omega_c$.
Such a transmission line cavity can be modeled around its fundamental mode resonance by a lumped element RLC circuit with the equivalent capacitor, inductor and resistor
\begin{equation}
C = \frac{C'l}{2}, ~~~~~ L = \frac{8}{\pi^2}L'l, ~~~~~ R = Z_0\alpha l
\end{equation}
respectively.
Here, $C'$ and $L'$ denote capacitance and inductance of the transmission line per unit length and $\alpha$ is the line attenuation constant.
For a capacitively coupled parallel RLC circuit, the ideal $S_{21}$ response function is in high-$Q$ approximation given by
\begin{equation}
S_{21} = 1 - \frac{\kappa_e}{\kappa_i + \kappa_e + 2i\Delta}
\label{eqn:idealS}
\end{equation}
with the internal and external decay rates
\begin{equation}
\kappa_i = \frac{1}{R(C + C_c)}, ~~~~~ \kappa_e = \frac{\omega_c^2 C_c^2 Z_0}{2(C + C_c)}
\end{equation}
and the detuning from the resonance frequency
\begin{equation}
\Delta = \omega - \omega_c, ~~~~~ \omega_c = \frac{1}{\sqrt{L(C+C_c)}}.
\end{equation}

\subsection{Extracting cavity parameters from data}

In any real experiment with microwave cables and microwave elements such as attenuators, circulators and amplifiers, the measured resonance line is not described by Eq.~(\ref{eqn:idealS}) anymore.
To model the measured complex scattering parameter $S_{21}$, we use
\begin{equation}
S_{21} = (\alpha_0 + \alpha_1\omega) \left(1 - \frac{\kappa_e e^{i\theta}}{\kappa_i + \kappa_e +2i\Delta}\right) e^{i{(\beta_1 \omega  + \beta_0)}}
\label{eqn:realS}
\end{equation}
where we consider a modification of the background signal and phase by using the frequency dependent complex scaling factor
\begin{equation}
(\alpha_0 + \alpha_1\omega)\cdot e^{i{(\beta_1 \omega  + \beta_0)}}
\end{equation}
and also include an additional rotation of the complex resonance circle around its anchor point by the phase factor $e^{i\theta}$.
\begin{figure}[h]
	\centering {\includegraphics[trim={2cm 20.5cm 2cm 2cm},clip=True,scale=0.85]{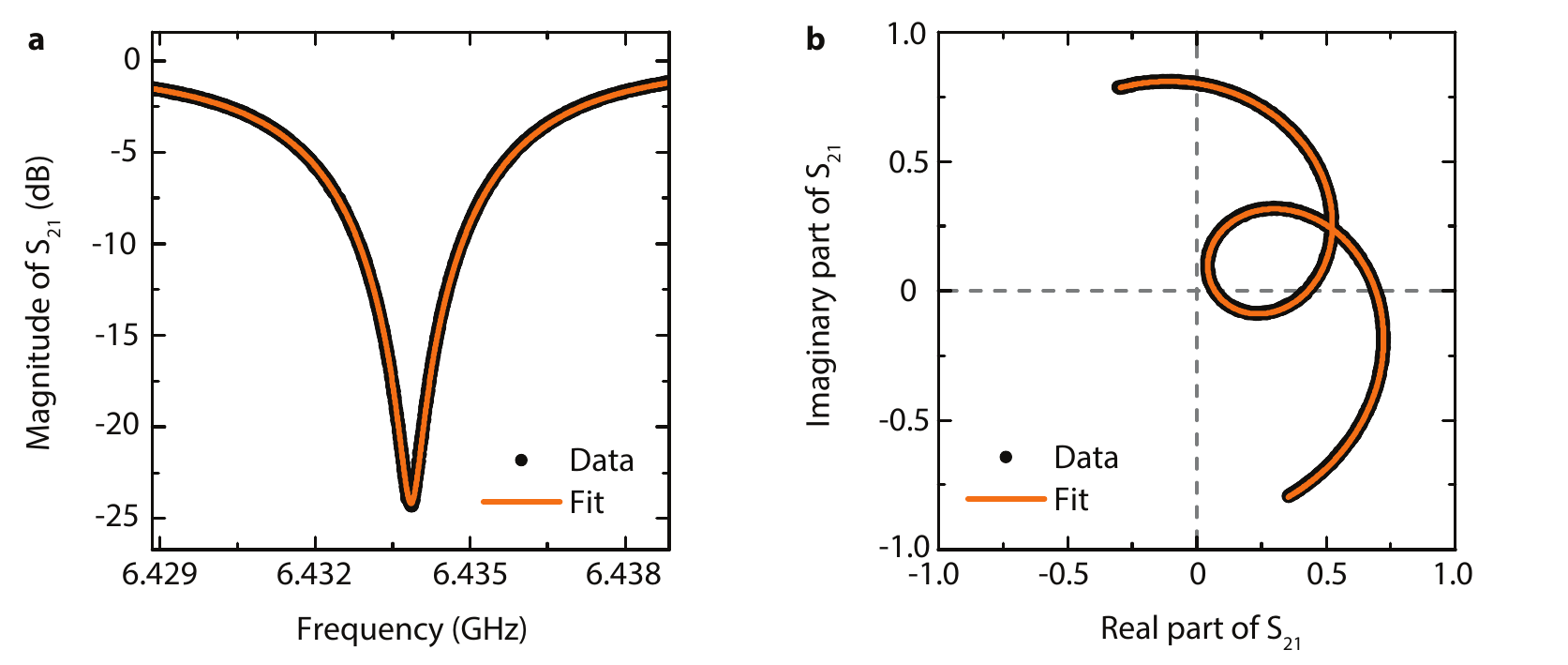}}
	\caption{\textsf{\textbf{Fitting the resonance line and extraction of the relevant parameters.} \textbf{a} shows the magnitude of $S_{21}$ and \textbf{b} the response in the complex plane. In both panels, data are shown as black circles and the fit as orange line.}}
	\label{fig:Fit}
\end{figure}

Figure~\ref{fig:Fit} shows an experimentally determined resonance curve in both, magnitude (\textbf{a}) and the complex plane (\textbf{b}), in direct comparison with the fit we obtained using Eq.~(\ref{eqn:realS}).
Both curves are normalized by $\alpha_0 + \alpha_1\omega_c$, i.e., by the background value at the fitted resonance frequency.
From the fit, we extract the cavity parameters $\kappa_i = 2\pi\cdot 370\,$kHz, $\kappa_e = 2\pi\cdot 5.7\,$MHz, and $\omega_c = 2\pi\cdot 6.4339\,$GHz.
Thus, the cavity is highly overcoupled with a coupling efficiency $\eta = \kappa_e/(\kappa_i + \kappa_e) = 0.94$.

\subsection{Cavity parameters vs sideband drive power}

\begin{figure}[h]
	\centering {\includegraphics[trim={1.cm 19.5cm 1.cm 4.cm},clip=True,scale=0.85]{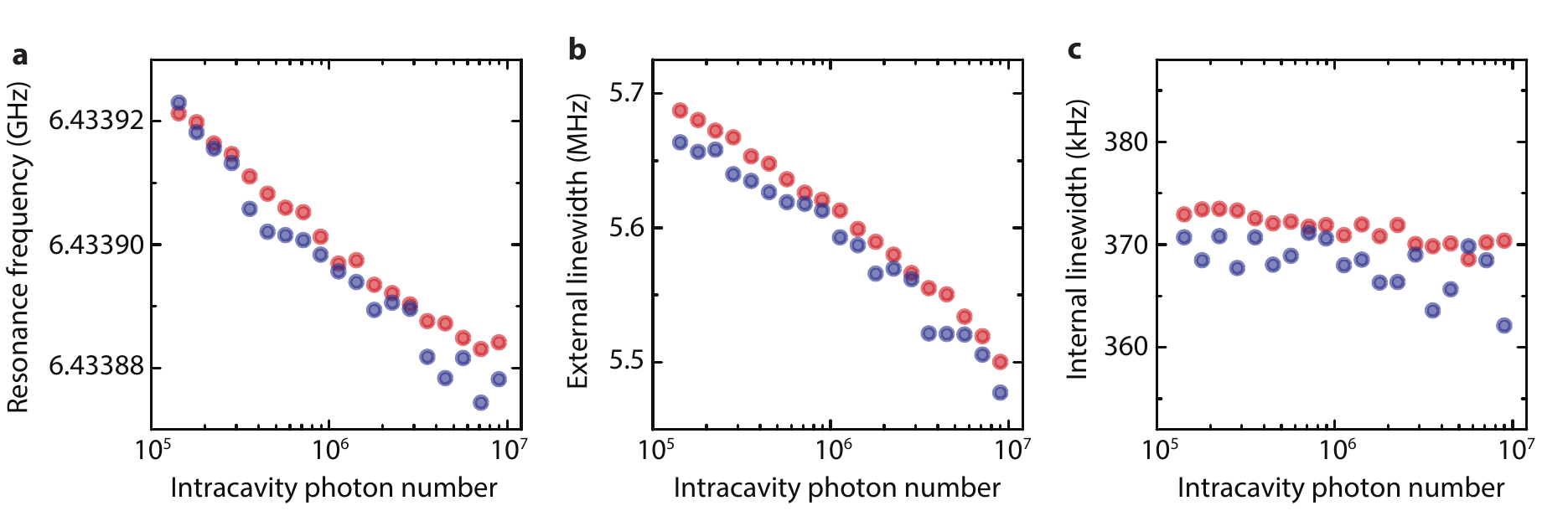}}
	\caption{\textsf{\textbf{Cavity parameters in presence of a sideband drive vs intracavity photon number.} Red data points correspond to a drive at $\omega \sim \omega_c - \Omega_m$, blue data points to a drive at $\omega \sim \omega_c + \Omega_m$.}}
	\label{fig:CavitySBD}
\end{figure}

In the parts of the experiments, where we investigate optomechanically induced transparency and demonstrate microwave amplification, we add a high-power microwave tone on one of the cavity sidebands.
This strong tone slightly modifies the cavity linewidths and the cavity resonance frequency depending on its power.
In Fig.~\ref{fig:CavitySBD}, we plot the resonance frequency (\textbf{a}) as well as external (\textbf{b}) and internal (\textbf{c}) cavity linewidths as extracted from fitting the corresponding curves with Eq.~(\ref{eqn:realS}).
The frequency of the sideband drive was set to $\omega \approx \omega_c \pm \Omega_m$ and kept fixed for all powers and the probe tone power was much smaller than the sideband drive.
Thus, due to the slight dependence of the cavity resonance frequency on the drive power it is not exactly on the red/blue sideband for all powers.
The deviation from the low-power resonance frequency, however, is given by maximally $0.007(\kappa_i + \kappa_e)$, i.e., less than one percent of the linewidth, and thus we consider it as negligible.

The photon numbers in the cavity are calculated by using
\begin{equation}
n = \frac{2P_\mathrm{in}}{\hbar \omega_d}\frac{\kappa_e}{\kappa^2 + 4\Omega_m^2},
\end{equation}
where $P_\mathrm{in}$ is the input power on the chip feedline, $\omega_d$ is the drive frequency and we assume a detuning of the drive tone from the cavity resonance by one mechanical frequency $\Delta = -\Omega_m$.

\subsection{Coupling capacitance and characteristic impedance}

With the formula for the external decay rate for a capacitively side-coupled RLC circuit
\begin{equation}
\kappa_e = \frac{\pi\omega_c C_c^2}{8\sqrt{C(C+C_c)^3}}
\end{equation}
the calculated capacitance per unit length $C' = 187\,$pF/m for our coplanar waveguide geometry (center conductor width $S = 10\,\mu$m, gap width $W = 6\,\mu$m, substrate permittivity $\epsilon_r = 11.6$) and the cavity length $l = 3450\,\mu$m, we can determine the coupling capacitance as $C_c \approx 16\,$fF.
In the expression for $\kappa_e$, we have used that the resonance impedance of the equivalent RLC circuit $Z_r = \sqrt{L/C} = \frac{4}{\pi}Z_0$ is related to the feedline impedance $Z_0$ for our geometry.
With the value for the coupling capacitance, we calculate the equivalent inductance $L = 1.8\,$nH, the inductance per unit length $L' = 645\,$nH/m, which is considerably larger than the calculated geometric inductance per unit length $L_g' = 375\,$nH/m due to kinetic contributions, and finally the characteristic impedance of feedline and cavity as $Z_0 = 65.6\,\Omega$.

\section{Theory of optomechanical motion detection}

When a microwave signal is sent into the cavity on resonance, the ideal response is given by
\begin{equation}
V(t) = V_\mathrm{\omega}\left(\frac{\kappa_i}{\kappa_i + \kappa_e}\right)e^{i\omega t}.
\end{equation}
If on the other hand the resonance frequency is modulated by mechanical motion, i.e., $\omega_c = \omega_c - G x(t)$ with the cavity pull $G = -\frac{\partial \omega_c}{\partial x}$ and under the assumptions that $x(t)$ is a real-valued function and neglecting small transients $(\kappa/\Omega_m \approx 4)$, we get
\begin{eqnarray}
V(t) & = & V_\mathrm{\omega}\left(\frac{\kappa_i + 2iGx(t)}{\kappa_i + \kappa_e + 2iGx(t)}\right)e^{i\omega t}\\
 & \approx & V_\mathrm{\omega}\left( \frac{\kappa_i}{\kappa_i + \kappa_e} + 2i G\frac{\kappa_e}{(\kappa_i + \kappa_e)^2}x(t)\right)e^{i\omega t}.
\end{eqnarray}
where the approximation in the last step was done for $G^2x^2 \ll (\kappa_i + \kappa_e)^2$, i.e., the motion induced frequency shift is much smaller than the cavity linewidth.
Now assuming that the mechanical position is given by
\begin{equation}
x(t) = x_0 \cos{\Omega t} = \frac{x_0}{2}\left(e^{i\Omega t} + e^{-i\Omega t}\right)
\end{equation}
we get for the response voltage
\begin{equation}
V(t) = V_\omega\frac{\kappa_i}{\kappa}e^{i\omega t} + iV_\omega Gx_0\frac{\kappa_e}{\kappa^2}\left(e^{i(\omega + \Omega)t} + e^{i(\omega - \Omega)t}\right).
\end{equation}
To calculate the effect of mixing this response with a signal oscillating with $\omega$, as we do in the experiment, we take the real part first given by
\begin{equation}
V_r(t) = V_\omega\frac{\kappa_i}{\kappa}\cos{\omega t} - V_\omega G x_0 \frac{\kappa_e}{\kappa^2}\left[\sin{(\omega + \Omega)t + \sin{(\omega - \Omega)t}} \right]
\end{equation}
and multiply this with a mixing local oscillator $\cos{(\omega t + \gamma)}$ containing an arbitrary phase offset $\gamma$.
The result is given by
\begin{equation}
V_f(t) = \frac{V_\omega}{2}\frac{\kappa_i}{\kappa}\cos{\gamma} - \frac{V_\omega}{2} \frac{\kappa_e}{\kappa^2}Gx_0\cos{\Omega t}\sin{\gamma} + ...
\end{equation}
where we omitted frequency components oscillating with $2\omega$ or $2\omega \pm \Omega$.
Thus, if the mixer phase $\gamma$ is different from zero or $\pi$, this technique will generate a signal with the frequency of the mechanical motion and the amplitude of this signal is proportional to the mechanical displacement amplitude.
This way we detected both the mechanical amplitude amplification as well as the thermal noise squeezing in this work.

\section{Nanowire characterization}

\subsection{Nanowire tuning with a DC voltage}

In this paper, we describe the nanowire as a point-like mechanical harmonic oscillator, cf. Sec.~\ref{sec:TPMA}.
When a DC voltage is applied to the center conductor of the transmission feedline, cf. Fig.~\ref{fig:Setup}\textbf{a} and main paper Fig.~1, a static force is exerted to the nanowire and the equation of motion is given by
\begin{equation}
m\ddot{x} + m\Gamma_m \dot{x} + k_mx = \frac{1}{2}V_\mathrm{dc}^2\frac{\partial C_\mathrm{nw}}{\partial x},
\end{equation}
were $x$ is the nanowire position, $k_m$ is the intrinsic spring constant, $m$ is the effective mass and $\Gamma_m$ is the intrinsic damping or mechanical linewidth.
The force will lead to a new nanowire equilibrium position $x_0$ which is defined by
\begin{equation}
k_m x_0 = \frac{1}{2}V_\mathrm{dc}^2\frac{\partial C_\mathrm{nw}}{\partial x}\bigg|_{x_0}.
\end{equation}
A Taylor approximation of the electrostatic force around the new equilibrium position $x_0$ gives
\begin{equation}
F_\mathrm{el} = \frac{1}{2}V_\mathrm{dc}^2\left[\frac{\partial C_\mathrm{nw}}{\partial x}\bigg|_{x_0} + \frac{\partial^2 C_\mathrm{nw}}{\partial x^2}\bigg|_{x_0}(x - x_0) ...\right].
\end{equation}
Absorbing the new equilibrium position in a redefinition of the position coordinate $x$ allows to write the full equation of motion as
\begin{equation}
m\ddot{x} + m\Gamma_m\dot{x} + \left[k_m + k_\mathrm{dc}\right]x = F(t)
\end{equation}
where
\begin{equation}
k_\mathrm{dc} = -\frac{1}{2}V_\mathrm{dc}^2\frac{\partial^2 C_\mathrm{nw}}{\partial x^2}\bigg|_{x_0}
\end{equation}
is the electrostatic spring constant and $F(t)$ is a possible additional external driving force.
In general, the equilibrium position and the second derivative of the capacitance will depend on the applied DC voltage themselves.
From the equation of motion it follows that the mechanical resonance frequency is given by
\begin{eqnarray}
\Omega_m & = & \sqrt{\frac{k_m + k_\mathrm{dc}}{m}}\\
& = &\Omega_{m0}\sqrt{1 + \frac{k_\mathrm{dc}}{k_m}}
\label{eqn:Tuning}
\end{eqnarray}
where $\Omega_{m0} = \sqrt{k_m/m}$ is the intrinsic mechanical resonance frequency.
Note that the electrostatic spring constant is negative and that the resonance frequency is shifted to lower values.
To characterize the mechanical oscillator, we drive it with an additional near-resonant harmonic voltage as described in Sec.~\ref{sec:TPMA} and measure the resonance peak for different DC voltages.
With an estimate for the mass of the nanobeam, we can extract the effective spring constant from the zero voltage resonance frequency.
With the dimensions of the beam, its full mass is calculated by using the densities $\rho_\mathrm{SiN} = 3.2\,$g/cm$^3$ and $\rho_\mathrm{MoRe} = 14.5\,$g/cm$^3$ to be $m = 17\,$pg.

\begin{figure}[h]
	\centering {\includegraphics[trim={1.cm 19.5cm 1.cm 1.5cm},clip=True,scale=0.7]{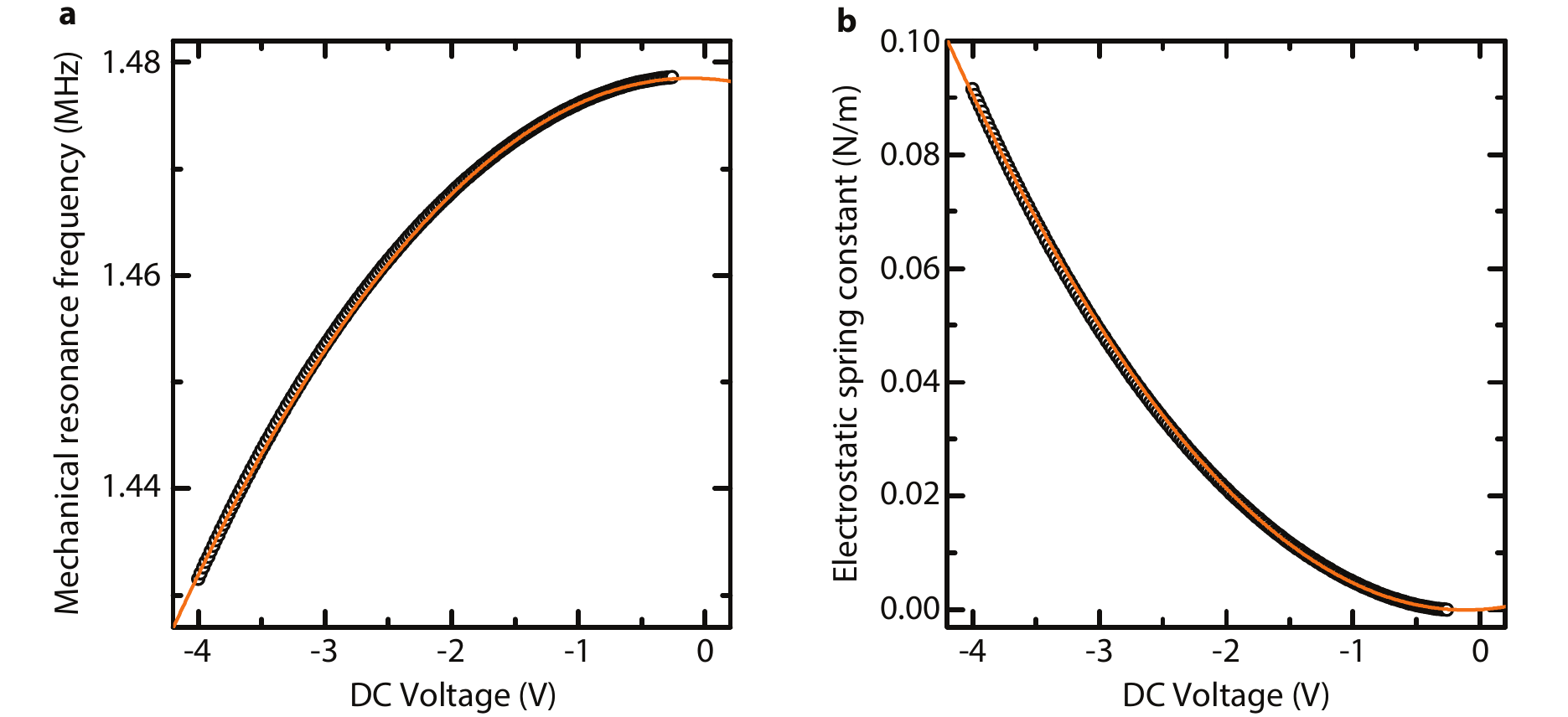}}
	\caption{\textsf{\textbf{Mechanical resonance frequency tuning and electrostatic spring constant.} In \textbf{a} the measured resonance frequency is plotted vs applied DC voltage on the feedline (black circles). The orange line is a fit using Eq.~(\ref{eqn:Tuning}) with $k_\mathrm{dc} \propto V_\mathrm{dc}^2$. In \textbf{b} the magnitude of the electrostatic spring constant calculated from the fit in \textbf{a} is shown as orange line. The experimental data (black circles) are calculated using Eq.~(\ref{eqn:elSpring}). At the operation point of this paper $V_\mathrm{dc} = -4\,$V, we obtain $k_\mathrm{dc} \approx - 0.09\,$N/m.}}
	\label{fig:Tuning}
\end{figure}

From the resonance frequency $\Omega_{m0} = 2\pi\cdot1.478\,$MHz, we can thus extract the effective intrinsic spring constant $k_m = 1.46\,$N/m.
In addition, we can calculate the electrical spring constant $k_\mathrm{dc}$ from here.
The measured resonance frequency vs DC voltage in the negative voltage range is plotted in Fig.~\ref{fig:Tuning}\textbf{a} and a fit with Eq.~(\ref{eqn:Tuning}) assuming $k_\mathrm{dc} \propto V_\mathrm{dc}^2$ describes the behaviour very accurately in the shown voltage range.
With Eq.~(\ref{eqn:Tuning}) we can also calculate the electrostatic spring constant
\begin{equation}
k_\mathrm{dc} = k_m\left(1 - \frac{\Omega_m^2}{\Omega_{m0}^2}\right)
\label{eqn:elSpring}
\end{equation}
from our data and $k_m = 1\,$N/m.
The result is plotted together in Fig.~\ref{fig:Tuning}\textbf{b} with a line obtained from the fit in \textbf{a}.

\section{Optomechanical device characterization}

\subsection{Optomechanical coupling rate $g_0$}

To calculate the optomechanical single-photon coupling rate
\begin{equation}
g_0 = -\frac{\partial \omega_c}{\partial x} x_\mathrm{zpf}
\end{equation}
we need the mechanical zero-point fluctuations, which we get from the resonance frequency and the effective nanowire mass as
\begin{equation}
x_\mathrm{zpf} = \sqrt{\frac{\hbar}{2 m \Omega_m }} = 18\,\mathrm{fm}.
\end{equation}
To get the cavity pull parameter, we estimate from simulations and calculations the mechanical capacitance to be approximately $C_m = 2\,$fF and calculate
\begin{equation}
\frac{\partial \omega_c}{\partial C_m} = -\frac{\omega_c^3}{2} L = -2\pi\cdot 9.5\cdot 10^{21}\,\mathrm{Hz/F}.
\end{equation}
The final quantity we need is
\begin{equation}
\frac{\partial C_m}{\partial x} \approx 5 \cdot 10^{-9}\,\mathrm{F/m},
\end{equation}
which gives a cavity pull of
\begin{equation}
G = -\frac{\partial\omega_c}{\partial x} = 2\pi\cdot 48\,\mathrm{kHz/nm}.
\end{equation}
Putting everything together we get
\begin{equation}
g_0 = 2\pi\cdot 0.9\,\mathrm{Hz}.
\end{equation}

\subsection{Optomechanical coupling ratio}

Resonance frequency and external coupling rate of our cavity are given by
\begin{eqnarray}
\omega_c & = & \frac{1}{\sqrt{L(C+C_c)}}\\
\kappa_e & = & \frac{Z_0 C_c^2}{2L(C+C_c)^2}
\end{eqnarray}
with the characteristic feedline impedance $Z_0$. 
Assuming that the cavity is highly overcoupled $\kappa_e + \kappa_i \approx \kappa_e$ as in our device, the ratio of dissipative optomechanical coupling rate $g_\kappa$ to dispersive optomechanical  coupling rate $g_\omega$ is given by
\begin{equation}
\frac{g_\kappa}{g_\omega} = 2Z_0\omega_c \frac{C C_c}{C + C_c} \approx 0.08,
\end{equation}
which is small enough to neglect the dissipative optomechanical coupling contribution to first order throughout the paper.
Thus, we will restrict our theoretical calculations and device modeling to purely dispersive coupling.

\subsection{Theory of optomechanically induced transparency and absorption without the resolved sideband limit}

We model the system without the parametric driving by means of the classical, coupled equations of motion for the mechanical displacement $x$ and the intracavity field amplitude $\alpha$
\begin{eqnarray}
\ddot{x} = -\Omega_m^2 x - \Gamma_m \dot{x} + \frac{1}{m}\left(F_r + F_e\right)\\
\dot{\alpha} = \left[i(\Delta + G x) - \frac{\kappa}{2}\right]\alpha + \sqrt{\frac{\kappa_e}{2}}S_\mathrm{in}
\end{eqnarray}
where external forces to the mechanical oscillator are expressed by $F_e$ in the first equation and the radiation pressure force due to the intracavity field is given by
\begin{equation}
F_r = \hbar G |\alpha|^2.
\end{equation}

Further parameters in the equations are the cavity pull parameter $G = -\partial \omega_c/\partial x$, the detuning between a cavity drive and the cavity resonance frequency $\Delta = \omega_d - \omega_c$ and the total cavity linewidth $\kappa = \kappa_i + \kappa_e$.
In the second equation, the field amplitude $\alpha$ is normalized such that $|\alpha|^2$ corresponds to the photon number in the cavity and the input field $S_\mathrm{in}$ is normalized such that $|S_\mathrm{in}|^2$ corresponds to the photon number flux of the input field.
Under the assumption that it is sufficient to consider only small deviations from the steady state solutions $\bar{x}, \bar{\alpha}$ of the full equations, i.e., $x = \bar{x} + \delta x, \alpha = \bar{\alpha} + \delta\alpha$, these two equations can be linearized as
\begin{eqnarray}
\delta\ddot{x} & = & -\Omega_m^2 \delta x - \Gamma_m \delta\dot{x} + \frac{\hbar G \bar{\alpha}}{m}\left(\delta\alpha + \delta\alpha^*\right) \\
\delta\dot{\alpha} & = & \left[i\bar{\Delta} - \frac{\kappa}{2}\right]\delta\alpha + iG\bar{\alpha}\delta x + \sqrt{\frac{\kappa_e}{2}}S_p.
\end{eqnarray}
where we omitted a possible external driving force $F_e$.
Here, $\bar{\Delta} = \omega_d - \omega_c + G\bar{x}$ is the detuning from the modified resonance frequency, when the mechanical oscillator is pushed by radiation pressure to its new equilibrium position $\bar{x}$, and $\sqrt{\kappa_e/2}S_p$ with $S_p = S_0e^{-i\Omega t}$ ($\Omega = \omega - \omega_d$) accounts for small additional drive fields or field fluctuations.
We solve these equations with the Ansatz
\begin{eqnarray}
\delta\alpha & = & a_-e^{-i\Omega t} + a_+e^{+i\Omega t}\\
\delta\alpha^* & = & a^*_-e^{+i\Omega t} + a_+^*e^{-i\Omega t}\\
\delta x & = & x_1e^{-i\Omega t} + x_1^* e^{+i\Omega t}
\end{eqnarray}
and get as solution in high-$Q_m$ approximation the modified mechanical response function
\begin{equation}
\chi_m^\mathrm{eff} = \frac{1}{2m\Omega_m}\frac{1}{\Omega_m - \Omega - i\frac{\Gamma_m}{2} + \Sigma'(\Omega_m)}
\end{equation}
where
\begin{equation}
\Sigma'(\Omega_m) = -ig^2\left[\chi_c(\Omega_m) - \chi_c^*(-\Omega_m)\right].
\label{eqn:Sigma}
\end{equation}
Here,
\begin{equation}
\chi_c = \frac{1}{\frac{\kappa}{2} - i(\bar{\Delta} + \Omega)}
\end{equation}
with $\bar{\Delta} = \omega_d - \omega_c + G\bar{x}$ represents essentially the (modified) cavity response lineshape, for which we use from here on just $\Delta$ as the difference is negligibly small in our experiment.
Expression (\ref{eqn:Sigma}) can be split into an imaginary and a real part $\Sigma' = \delta\Omega_m - i\Gamma_o/2$, of which the real part corresponds to a modification of the mechanical resonance frequency (optical spring)
\begin{equation}
\delta\Omega_m = g^2\left[\frac{\Delta + \Omega_m}{\frac{\kappa^2}{4} + (\Delta + \Omega_m)^2} + \frac{\Delta - \Omega_m}{\frac{\kappa^2}{4} + (\Delta - \Omega_m)^2}\right]
\end{equation}
and the imaginary part
\begin{equation}
\Gamma_o = g^2\kappa\left[\frac{1}{\frac{\kappa^2}{4} + (\Delta + \Omega_m)^2} - \frac{1}{\frac{\kappa^2}{4} + (\Delta - \Omega_m)^2}\right]
\end{equation}
represents an additional damping term (optical damping).
For the cavity amplitude, we find the solution
\begin{equation}
a_- = \chi_c\left[1 + 2im\Omega_m g^2 \chi_c\chi_m^\mathrm{eff}\right]\sqrt{\frac{\kappa_e}{2}}S_0
\end{equation}
which with $S_{21} = 1-\sqrt{\frac{\kappa_e}{2}}\frac{a_-}{S_0}$ can be directly translated into the full cavity response function in presence of a harmonic drive
\begin{equation}
S_{21} = 1-\frac{\kappa_e}{2}\chi_c\left[1 + 2im\Omega_m g^2 \chi_c \chi_m^\mathrm{eff}\right].
\end{equation}

\subsubsection{Drive on the red sideband}

When the constant frequency drive is set to the red cavity sideband, i.e., $\Delta = -\Omega_m$, and the probe tone is swept only very close to the cavity resonance, i.e., $\Omega = \Omega_m + \Delta_m$  with $\Delta_m \ll \kappa$, the effective cavity susceptibility is given by
\begin{equation}
\chi_c = \frac{2}{\kappa}
\end{equation}
and the effective mechanical susceptibility can be approximated as
\begin{equation}
\chi_m^\mathrm{eff} = -\frac{1}{m\Omega_m}\frac{1}{2\Delta_m + i\Gamma_\mathrm{eff}}
\end{equation}
with
\begin{equation}
\Gamma_\mathrm{eff} = \Gamma_m + \Gamma_o = \Gamma_m\left(1 + C \frac{16\frac{\Omega_m^2}{\kappa^2}}{1 + 16\frac{\Omega_m^2}{\kappa^2}}\right)
\end{equation}
where $C = 4g^2/\kappa\Gamma_m$ is the cooperativity.
The scattering parameter is then given by
\begin{eqnarray}
S_{21} & = & 1 - \frac{\kappa_e}{\kappa}\left[1 - 4i\frac{g^2}{\kappa}\frac{1}{2\Delta_m + i\Gamma_\mathrm{eff}}\right]\\
& = & \frac{\kappa_i}{\kappa} + i\frac{\kappa_e}{\kappa}\frac{C\Gamma_m}{2\Delta_m + i\Gamma_\mathrm{eff}}.
\end{eqnarray}
The transmitted power is then described by a Lorentzian
\begin{equation}
|S_{21}|^2 = S_c + \frac{C\Gamma_m}{4\Delta_m^2 + \Gamma_\mathrm{eff}^2}S_\mathrm{om}
\label{eqn:S21OMIT}
\end{equation}
with the background value
\begin{equation}
S_c = \frac{\kappa_i^2}{\kappa^2}
\end{equation}
and the optomechanical amplitude
\begin{equation}
S_\mathrm{om} = 2\frac{\kappa_i \kappa_e}{\kappa^2}\Gamma_\mathrm{eff} + \frac{\kappa_e^2}{\kappa^2}C \Gamma_m.
\end{equation}

\subsubsection{Drive on the blue sideband}

With $\Delta = +\Omega_m$ and $\Omega \approx -\Omega_m + \Delta_m$ we get

\begin{equation}
\chi_m^\mathrm{eff} = \frac{1}{m\Omega_m}\frac{1}{2\Delta_m + i\Gamma_\mathrm{eff}'}
\end{equation}

where

\begin{equation}
\Gamma_\mathrm{eff}' = \Gamma_m - \Gamma_o = \Gamma_m\left(1 - C \frac{16\frac{\Omega_m^2}{\kappa^2}}{1 + 16\frac{\Omega_m^2}{\kappa^2}}\right)
\end{equation}

As transmission parameter we thus get
\begin{eqnarray}
S_{21} & = & 1 - \frac{\kappa_e}{\kappa}\left[1 + 4i\frac{g^2}{\kappa}\frac{1}{2\Delta_m + i\Gamma_\mathrm{eff}'}\right]\\
& = & \frac{\kappa_i}{\kappa} - i\frac{\kappa_e}{\kappa}\frac{C \Gamma_m}{2\Delta_m + i\Gamma_\mathrm{eff}'}
\end{eqnarray}
and for the transmitted power
\begin{equation}
|S_{21}|^2 = S_c + \frac{C\Gamma_m}{4\Delta_m^2 + \Gamma_\mathrm{eff}'^2}S_\mathrm{om}'
\label{eqn:S21OMIA}
\end{equation}
with
\begin{equation}
S_\mathrm{om}' = - 2\frac{\kappa_i \kappa_e}{\kappa^2}\Gamma_\mathrm{eff} + \frac{\kappa_e^2}{\kappa^2}C \Gamma_m
\end{equation}

\subsection{OMIT, OMIA and cooperativity with the device}

\begin{figure}[h]
	\centering {\includegraphics[trim={1.cm 2.cm 1.cm 1.5cm},clip=True,scale=1]{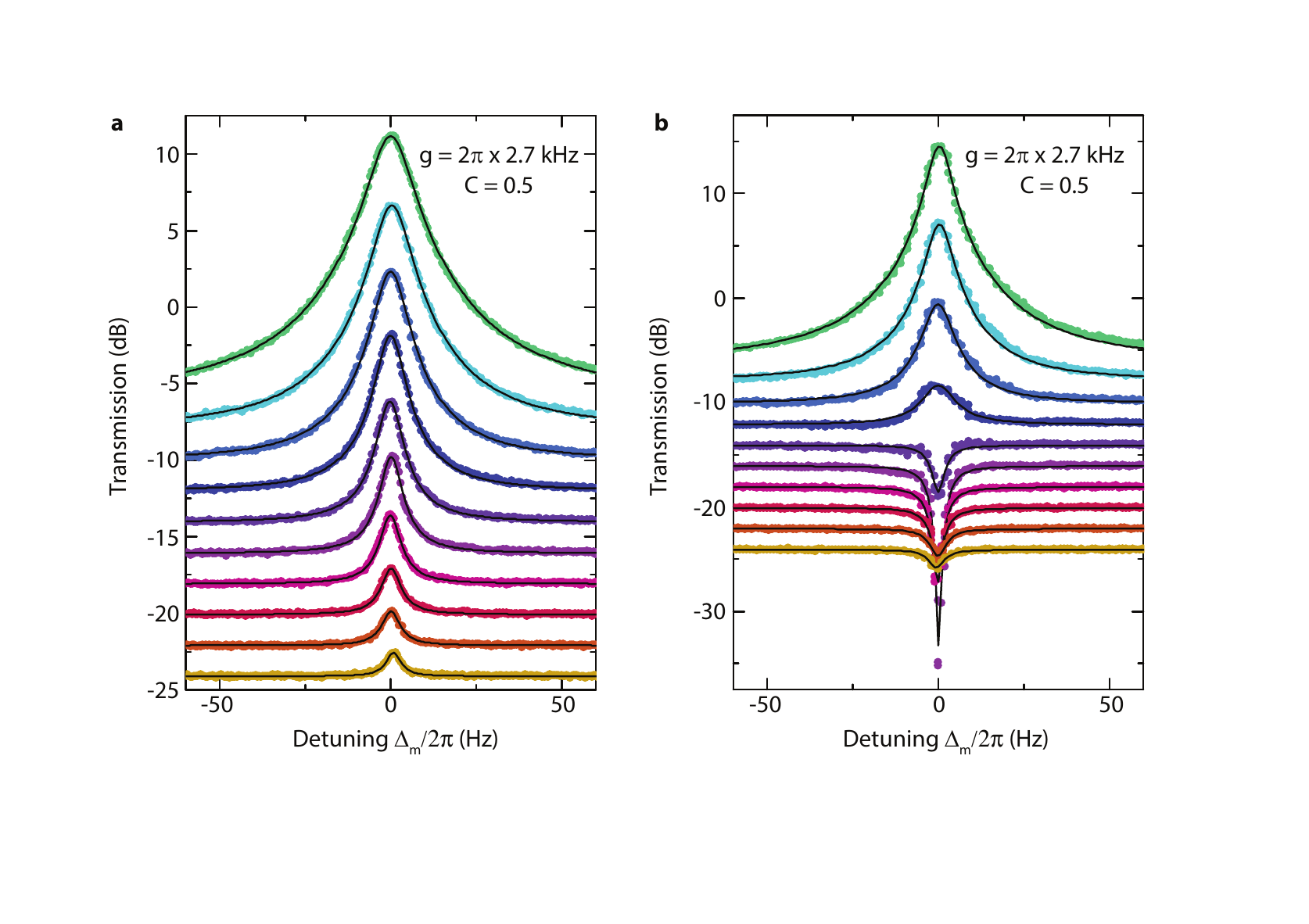}}
	\caption{\textsf{\textbf{Optomechanically induced transparency and absorption.} For this experiment we drive the cavity with a strong drive tone on either the red sideband $\omega = \omega_c - \Omega_m$ (\textbf{a}) or on the blue sideband $\omega = \omega_c + \Omega_m$ (\textbf{b}). Then we sweep a weak probe signal around $\Delta_m = \omega - \omega_c$. In \textbf{a}, a Lorentzian shaped peak corresponding to exciting the mechanical resonator appears and grows with increasing sideband drive power in both, height and width. In \textbf{b}, the cavity response for a drive on the blue sideband is shown. For the lower powers, optomechanically induced absorption appears, i.e., a narrow absorption dip in the cavity minimum. For higher powers, this dip turns into a transparency peak as well. The lowest line in both (baseline at $-24\,$dBm), \textbf{a} and \textbf{b}, corresponds to the lowest drive power and subsequent lines are manually upshifted by $2\,$dB each for better visibility. For both plots, the difference in drive power between subsequent lines is $2\,$dB, where the largest power corresponds to an intracavity drive photon number of $\sim 3\cdot 10^{5}$ for the red sideband and $\sim 3\cdot 10^{5}$ for the blue sideband drive.}}
	\label{fig:OMITOMIA}
\end{figure}

In the experiment, we drive the cavity with a drive tone on one of the sidebands, i.e. at $\omega = \omega_c \pm \Omega_m$ and variable power.
Then, we sweep a weak probe tone around the cavity resonance and measure the resulting optomechanically induced transparency or absorption.
Figure~\ref{fig:OMITOMIA} shows the resulting transparency and absorption windows for different drive powers, i.e., for different drive photon numbers inside the cavity.
In \textbf{a}, the optomechanically induced transparency window for a drive on the red sideband $\omega = \omega_c - \Omega_m$ is shown and in \textbf{b} the corresponding data for a drive on the blue sideband $\omega = \omega_c + \Omega_m$.
The different curves correspond to different drive powers (steps of $2\,$dBm) or intracavity photon numbers, respectively.

From the amplitude of the Lorentzians, we can extract the cooperativity and the total coupling rate $g$.
The extracted values for the highest power data are given in Fig.~\ref{fig:OMITOMIA}.
In combination with an estimate of the intracavity photon number, we get also 

\subsection{Effective mechanical linewidth}

\begin{figure}[h]
	\centering {\includegraphics[trim={0.cm 19.cm 0.cm 0.cm},clip=True,scale=0.65]{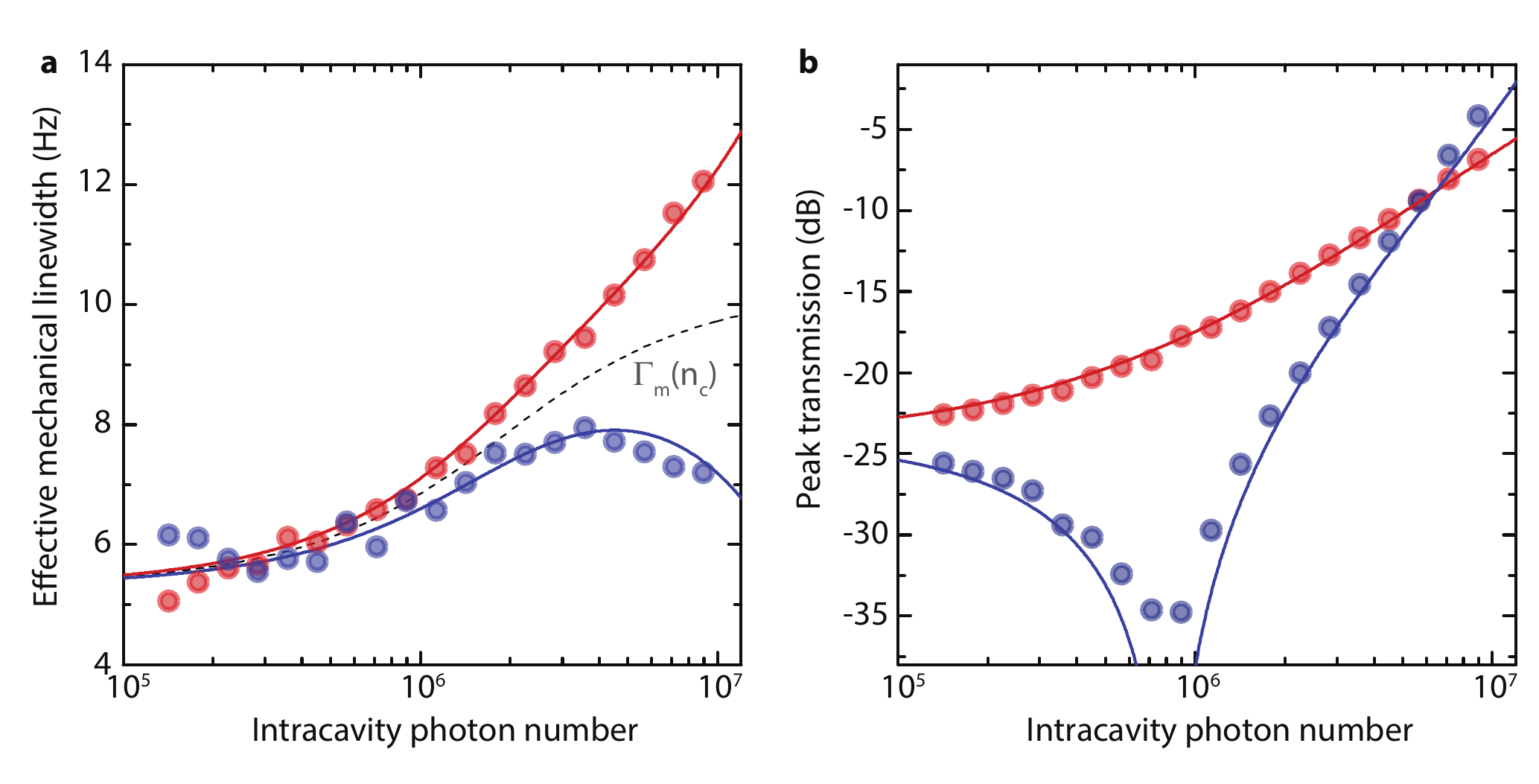}}
	\caption{\textsf{\textbf{Effective mechanical linewidth and OMIT peak transmission vs intracavity photon number.} \textbf{a} shows the effective mechanical linewidth extracted from the data in Fig.~\ref{fig:OMITOMIA} as points, red points for the red sideband, blue points for the blue. The dashed line shows the model for a power-dependent intrinsic linewidth $\Gamma_m(n_c)$ and the red and blue lines show the resulting calculated effective linewidth including dynamical backaction. In \textbf{b}, the peak transmission is plotted, circles correspond to experimental values and lines to theoretical calculations based on $\Gamma_m(n_c)$, $g_0$, $n_c$, $\kappa_i$, $\kappa_e$ and Eqs.~(\ref{eqn:S21OMIT}) and (\ref{eqn:S21OMIA}).}}
	\label{fig:Geff}
\end{figure}

The effective linewidth of the mechanical oscillator in an optomechanical system is given by the sum of the intrinsic linewidth $\Gamma_m$ and the optical linewidth $\Gamma_o$ due to dynamical backaction.
When extracting the linewidths from the data in Fig.~\ref{fig:OMITOMIA}, we find that we can best describe the overall dependence on the intracavity photon number by modelling a photon number dependent intrinsic linewidth $\Gamma_m(n_c)$, which might be caused by heating of the nanobeam, which is coupled directly to the feeedline.
When we model the intrinsic linewidth as shown by the dashed gray line in Fig.~\ref{fig:Geff}\textbf{a}, we find very good agreement between the theoretical lines of $\Gamma_\mathrm{eff} = \Gamma_m + \Gamma_o$ and the data points.
We use the phenomenological functional description $\Gamma_m(n_c) = \Gamma_{m0} + \gamma_1\arctan{\gamma_2n_c}$ here with $\gamma_i$ constant parameters adjusted to best describe the experimental data.
Calculating the peak transmission for this linewidth dependence in combination with the photon number, $g_0$, $\kappa_i$ and $\kappa_e$, gives excellent agreement between the theoretical curves and the data as shown in \textbf{b}.

\section{Theory of parametric mechanical amplitude amplification}
\label{sec:TPMA}

Similar to the description given in Ref.~\cite{Rugar91}, we model the nanowire as mechanical harmonic oscillator with the effective equation of motion of a point-like particle having the position coordinate $x$
\begin{equation}
m\ddot{x} + m\Gamma_m\dot{x} + k_mx = F(t).
\end{equation}
Here, $m$ is the effective mass of the nanowire, $\Gamma_m$ is its damping rate, $k_m$ is the mechanical spring constant and $F(t)$ is a time dependent external driving force.
When a time-dependent voltage $V(t)$ is applied to the center conductor of the microwave feedline, the nanowire experiences a corresponding electrical force
\begin{equation}
F_\mathrm{el} = \frac{1}{2}\frac{\partial C_\mathrm{nw}}{\partial x} V^2
\end{equation}
where $C_\mathrm{nw}$ is the capacitance between the nanowire and the center conductor of the cavity (DC ground).
In our experiment, the total voltage applied to the center conductor (without the microwave tone) is given by
\begin{equation}
V(t) = V_\mathrm{dc} + V_0\cos{\left(\Omega t + \phi_p\right)} + V_{2\Omega}\sin{2\Omega t}
\end{equation}
which corresponds to a total force of
\begin{eqnarray}
F_\mathrm{el}(t) & = \frac{1}{2}\frac{\partial C_\mathrm{nw}}{\partial x} & \Big[V_\mathrm{dc}^2 + V_0^2\cos^2{(\Omega t +\phi_p)} + V_{2\Omega}^2\sin^2{2\Omega t}\\
& & +2V_\mathrm{dc}V_0\cos{(\Omega t + \phi_p)} + 2V_\mathrm{dc}V_{2\Omega}\sin{2\Omega t}\\
& & +V_0V_{2\Omega}\cos{(\Omega t + \phi_p)}\sin{2\Omega t}\Big]
\end{eqnarray}
Here, $V_\mathrm{dc}$ is a static voltage, $V_\mathrm{0}$ is the voltage peak amplitude of a harmonic drive close to resonance with frequency $\Omega$ and $V_{2\Omega}$ is the corresponding amplitude of the parametric drive voltage with twice the frequency of the near-resonant drive.
We consider a phase shift between the near-resonant and the parametric drive by the phase $\phi_p$ in the near-resonant term.
In the experiment, we have used $V_\mathrm{dc} = -4\,$V, $V_{2\Omega}\leq 100\,\mu$V and $V_0 \approx 100\,$nV.
Keeping only the leading terms under these conditions, we get
\begin{equation}
F_\mathrm{el} \approx \frac{1}{2}\frac{\partial C_\mathrm{nw}}{\partial x} \left[V_\mathrm{dc}^2 + 2V_\mathrm{dc}V_0\cos{(\Omega t + \phi_p)} + 2V_\mathrm{dc}V_{2\Omega}\sin{2\Omega t}\right].
\end{equation}
A Taylor approximation to first order in $x$ around the equilibrium position $x_0$ gives
\begin{equation}
F_\mathrm{el} \approx \frac{1}{2}\left[V_\mathrm{dc}^2 + 2V_\mathrm{dc}V_0\cos{(\Omega t + \phi_p)} + 2V_\mathrm{dc}V_{2\Omega}\sin{2\Omega t}\right]\left[\frac{\partial C_\mathrm{nw}}{\partial x}\bigg|_{x_0} +  \frac{\partial^2 C_\mathrm{nw}}{\partial x^2}\bigg|_{x_0}\cdot(x-x_0) + ... \right]
\end{equation}
The first order terms proportional to $x-x_0$ can now be regarded as an electrostatic spring force with the spring constant
\begin{eqnarray}
k_\mathrm{el}(t) = \underbrace{-\frac{1}{2}\frac{\partial^2 C_\mathrm{nw}}{\partial x^2}\bigg|_{x_0}V_\mathrm{dc}^2}_{ = k_\mathrm{dc}}  \underbrace{-\frac{\partial^2 C_\mathrm{nw}}{\partial x^2}\bigg|_{x_0}V_\mathrm{dc}V_{2\Omega}}_{= k_p}\sin{2\Omega t}
\end{eqnarray}
where we have omitted the $\cos{(\Omega t + \phi_p)}$ term due to its smallness and the reduced effect of resonant parametric modulations compared to a $2\Omega$-term \cite{}.
Similarly, we can omit the $2\Omega$-term in the remaining driving force and after absorbing the remaining static force into a redefinition of the equilibrium position $x_0 = 0$ we finally get
\begin{equation}
m\ddot{x} + \Gamma_m \dot{x} + \left[k_0 + k_p\sin{2\Omega t}\right]x = F_0\cos{(\Omega t + \phi_p)}
\end{equation}
with 
\begin{equation}
k_0 = k_m + k_\mathrm{dc}, ~~~~~ F_0 = V_\mathrm{dc}V_0\frac{\partial C_\mathrm{nw}}{\partial x}\bigg|_{x_0}.
\end{equation}
This is the well-known equation of motion for a parametrically modulated harmonic oscillator \cite{Rugar91}.
With the transformations
\begin{equation}
\Omega_1 = \Omega_m\left[\left(1 - \frac{1}{4Q_m^2}\right)^{1/2} + \frac{i}{2Q_m}\right],
\end{equation}
\begin{equation}
A = \dot{x} + i\Omega_1^*x
\end{equation}
\begin{equation}
A^* = \dot{x} - i\Omega_1x
\end{equation}
we rewrite the equation of motion as
\begin{equation}
\dot{A} = i\Omega_1A + i\frac{k_p\sin{2\Omega t}}{m}\frac{A - A^*}{\Omega_1^* + \Omega_1} + \frac{F_0}{m}\cos{(\Omega t + \phi_p)}.
\end{equation}
With the Ansatz
\begin{equation}
A = A_0 e^{i\Omega t},
\end{equation}
and the high-$Q_m$ approximations
\begin{equation}
\Omega_1^* + \Omega_1 \approx 2\Omega_m,
\end{equation}
\begin{equation}
\Omega_1 - \Omega \approx i\frac{\Omega_m}{2Q_m} - \Delta_m
\end{equation}
where $\Delta_m = \Omega - \Omega_m$, we find in rotating wave approximation
\begin{equation}
\frac{\Omega_m}{2Q_m}A_0 + i\Delta_m A_0 + \frac{k_p}{4m\Omega_m}A_0^* - \frac{F_0}{2m}e^{i\phi_p} = 0.
\end{equation}
We can solve this equation for the real and the imaginary part of $A_0$. 
Setting $x(t) = x_1\cos{\Omega t} + x_2 \sin{\Omega t}$ and using $x_1 = \text{Im}(A_0)/\Omega_m$, $x_2 = \text{Re}(A_0)/\Omega_m$ we get 
\begin{eqnarray}
x_1 & = & F_0\frac{Q_m}{k_0}\left[\frac{\left(1 + \frac{Q_m k_p}{2k_0}\right)\sin{\phi_p} - \frac{2Q_m \Delta_m}{\Omega_m}\cos{\phi_p}}{1 - \frac{Q_m^2 k_p^2}{4 k_0^2} + \frac{4Q_m^2 \Delta_m^2}{\Omega_m^2}}\right]\\
x_2 & = & F_0\frac{Q_m}{k_0}\left[\frac{\left(1 - \frac{Q_m k_p}{2k_0}\right)\cos{\phi_p} + \frac{2Q_m \Delta_m}{\Omega_m}\sin{\phi_p}}{1 - \frac{Q_m^2 k_p^2}{4 k_0^2} + \frac{4Q_m^2 \Delta_m^2}{\Omega_m^2}}\right].
\end{eqnarray}
\begin{figure}[h]
	\centering {\includegraphics[trim={0cm 10cm 0cm 1cm},clip=True,scale=0.55]{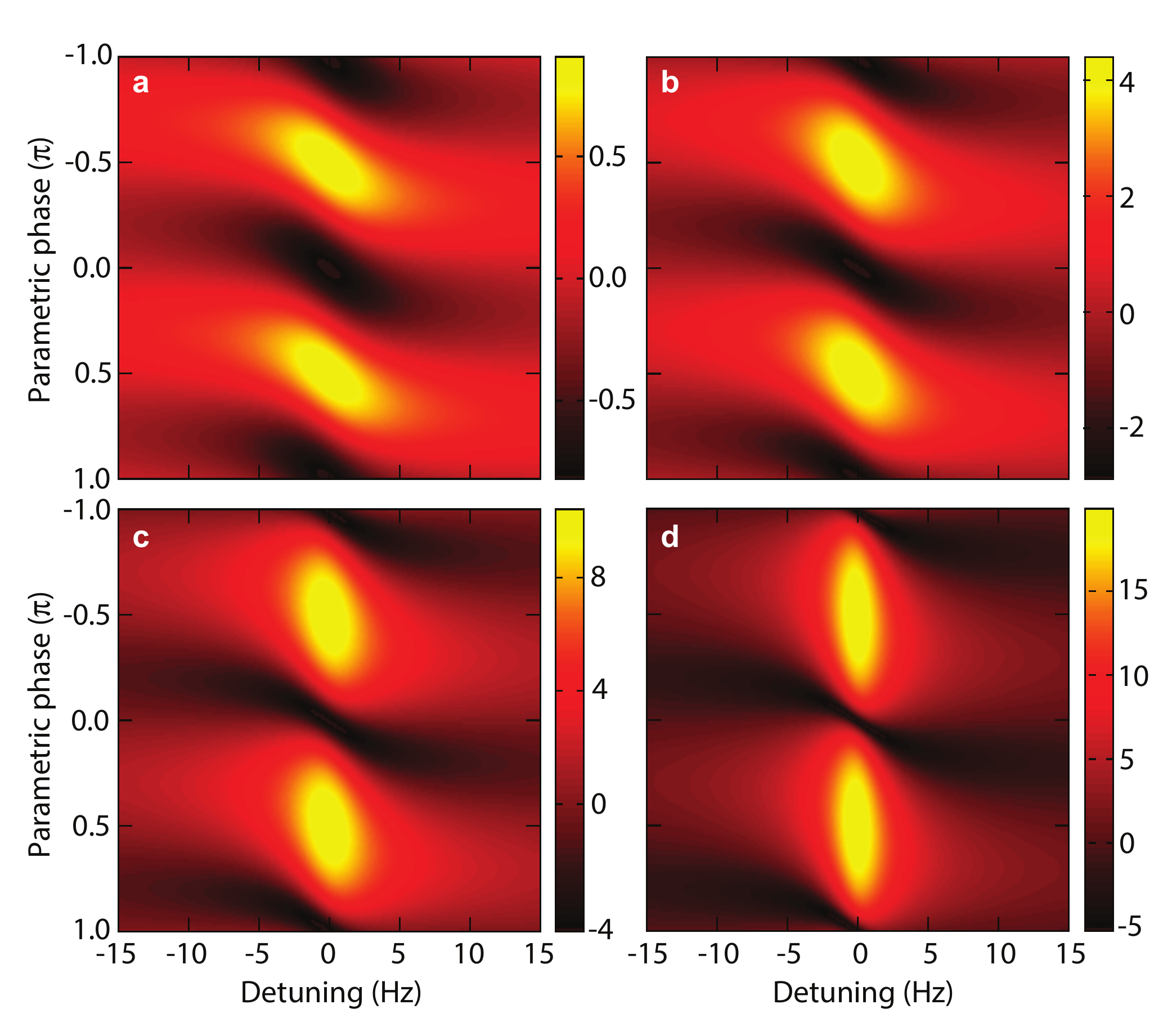}}
	\caption{\textsf{\textbf{Calculated mechanical parametric gain.} Plots show the mechanical parametric gain vs parametric phase $\phi_p$ and vs detuning from the mechanical resonance frequency $\Omega_m$ for four different parametric modulation amplitudes. \textbf{a} $V_{2\Omega}/V_t = 0.1$, \textbf{b} $V_{2\Omega}/V_t = 0.4$, \textbf{c} $V_{2\Omega}/V_t = 0.7$ and \textbf{d} $V_{2\Omega}/V_t = 0.9$ . The gain shows a $\pi$-periodicity and maxima/minima values follow an arctangent function with detuning. The calculation parameters were chosen close to the experimental device with $\Omega_m = 2\pi\cdot 1.4315\,$MHz and $Q_m = 195000$, the numbers at the color bars represent gain values in dB.}}
	\label{fig:MechGainTheory}
\end{figure}
From this we can calculate the mechanical amplitude as $|x| = \sqrt{x_1^2 + x_2^2}$ and get
\begin{equation}
|x|_\mathrm{on} = |x|_\mathrm{off}\left[\frac{\cos^2{(\phi_p + \varphi)}}{\left(1 + \frac{V_{2\Omega}}{V_t}\right)^2} + \frac{\sin^2{(\phi_p + \varphi)}}{\left(1 - \frac{V_{2\Omega}}{V_t}\right)^2}\right]^{1/2}
\label{eqn:MechAmp}
\end{equation}
where the amplitude without parametric drive is given by the usual expression
\begin{equation}
|x|_\mathrm{off} = F_0\frac{Q_m}{k_0}\frac{1}{\sqrt{1 + \frac{4Q_m^2 \Delta_m^2}{\Omega_m^2}}}
\end{equation}
describing the square root of a Lorentzian line around the resonance frequency.
Equation~(\ref{eqn:MechAmp}) is a generalization of the expression given in Ref.~\cite{Rugar91} for non-zero detunings from the resonance frequency.
The threshold voltage for parametric instability $V_t'$ in Eq.~(\ref{eqn:MechAmp}) describes the square root of an inverted Lorentzian
\begin{equation}
V_t = V_{t0}\sqrt{1 + \frac{4Q_m^2\Delta_m^2}{\Omega_m^2}},
\end{equation}
increasing with detuning from the resonant value
\begin{equation}
V_{t0} = \frac{2 k_0}{Q_m V_\mathrm{dc} \frac{\partial^2 C_\mathrm{nw}}{\partial x^2}},
\end{equation}
and the additional phase $\varphi$ appearing originates from the usual phase shift between the driving force and the mechanical response in absence of a parametric modulation.
It is given by
\begin{equation}
\varphi = -\frac{1}{2}\arctan{\left(\frac{2Q_m \Delta_m}{\Omega_m}\right)}.
\end{equation}
Finally, we can give the expression for the mechanical gain as
\begin{equation}
G = \frac{|x|_\mathrm{on}}{|x|_\mathrm{off}} = \left[\frac{\cos^2{(\phi_p + \varphi)}}{\left(1 + \frac{V_{2\Omega}}{V_t}\right)^2} + \frac{\sin^2{(\phi_p + \varphi)}}{\left(1 - \frac{V_{2\Omega}}{V_t}\right)^2}\right]^{1/2}
\label{eqn:MechGain}
\end{equation}
with the maximum and minimum values
\begin{equation}
G_\mathrm{max} = \frac{1}{1 - \frac{V_{2\Omega}}{V_t}}, ~~~~~ G_\mathrm{min} = \frac{1}{1 + \frac{V_{2\Omega}}{V_t}}
\end{equation}
for
\begin{equation}
\phi_{p}^\textrm{max} = \frac{\pi}{2}-\varphi, ~~~~~ \phi_{p}^\textrm{min} = -\varphi,
\end{equation}
respectively.
Figure~\ref{fig:MechGainTheory} shows the parametric mechanical gain as a function of parametric phase $\phi_p$ and detuning from the mechanical resonance frequency $\Delta_m = \Omega - \Omega_m$, calculated using Eq.~\ref{eqn:MechGain}.
The different panels show the amplitude gain for four different ratios of parametric drive voltage to threshold voltage $V_{2\Omega}/V_t$.
As expected from the equations, we find a $\pi$-periodicity of the gain and the maximum and minimum values follow an arctangent as a function of detuning from the resonance frequency.
We also see that with increasing parametric drive voltage and gain, respectively, the gradient of the gain with detuning also increases, demonstrating that frequency instabilities of the mechanical oscillator will lead to strong fluctuations of the gain as well in the high gain regime.

\section{Mechanical amplitude amplification - Measurement routine and data processing}

To measure the mechanical amplitude amplification, we sweep the phase between the drive tone and the parametric pump.
Instead of actually sweeping the phase, we added a small detuning on the order of $\sim 0.1\,$Hz to one of the tones, and we measured a time trace of the transmission signal at $\Omega$ while parametrically modulating the mechanical resonance frequency with $2\Omega+\delta$. 
Then, the parametric phase is given by $\phi_p = \delta t + \gamma$ with an arbitrary offset term $\gamma$.
We fitted the resulting power curves with
\begin{equation}
f(t) = \alpha_1\left[\frac{\cos^2(\alpha_2 t + \alpha_3)}{(1 + \alpha_4)^2} + \frac{\sin^2{\alpha_2 t + \alpha_3}}{(1-\alpha_4)^2}\right]
\label{eqn:GainFit}
\end{equation}
from where we get
\begin{equation}
G_\mathrm{min} = \frac{1}{1+\alpha_4}, ~~~~~~ G_\mathrm{max} = \frac{1}{1-\alpha_4}.
\end{equation}
Repeating this procedure for different detuning allows to determine the maximum and minimum gain dependent on $\Delta_m$.
Finally, we fit the detuning dependent maximum and minimum gain points with the corresponding theoretical expression
\begin{equation}
f(\Delta_m) = \frac{\beta_1}{\left(\sqrt{1 + \beta_2(\Delta_m - \beta_3)^2} \pm \beta_4\right)^2}
\end{equation}
where $\pm$ is chosen for minimum and maximum gain, respectively.
By this method, we can determine the maximum gain on resonance with higher fidelity than just setting the excitation frequency to the resonance frequency.
This is due to small mechanical resonance frequency drifts and fluctuations of unknown origin in the device.
Ultimately and for very high parametric excitation close to the threshold voltage, these frequency shifts also limit the observable gain, as it becomes more and more sensitive to frequency fluctuations as can also be seen in Fig.~\ref{fig:MechGainTheory}, where the range of largest gain gets narrower with increased $V_{2\Omega}/V_t$.

\section{Thermomechanical noise squeezing - Measurement routine and data processing}

\subsection{Measurement routine}
To characterize the thermomechanical noise of the nanobeam, we send a resonant microwave tone to the cavity and send the output field to a mixer with the drive tone as local oscillator.
This down-converts the motional sidebands to the original mechanical frequency.
To detect the quadratures of the sidebands $X'(t)$ and $Y'(t)$, we then measure the voltage with a lock-in amplifier set to the mechanical resonance frequency and a samplerate of $225\,$samples/s.
The total sampling time was $300\,$s.
This measurement scheme was repeated for different parametric modulation strengths, including without any parametric modulation, of the mechanical resonance frequency.
To characterize also the (white) background noise floor originating from our amplifier chain, we repeat the measurement for a lock-in center frequency sufficiently detuned from the mechanical resonance that there is no signature of the mechanical thermal noise included. 

\subsection{Data processing}

We calculate the power spectral density by PSD$ = |X'(\omega) + iY'(\omega)|^2$, where $X'(\omega)$ and $Y'(\omega)$ are the Fourier transforms of the recorded $X'(t)$ and $Y'(t)$.
The obtained spectra are smoothed by applying a simple $100$-point bin averaging and the result is shown in Fig.~3\textbf{a} of the main paper.
To calculate and plot the histograms, we first rotate the measured quadratures $X'(t)$ and $Y'(t)$ by $\sim \pi/36$ to obtain the amplified and de-amplified quadratures $X(t)$ and $Y(t)$.
Afterwards, we process the raw data by a $40$-point moving average and use every 5th of the resulting data points for the histograms.
We note, that the squeezing factor depends on the chosen data processing values for the averaging.

\section{Theory of parametric microwave amplification}

We include the parametric driving equivalently to the case of mechanical parametric amplification into the optomechanical equations of motion and get
\begin{eqnarray}
\delta\ddot{x} & = & -\left[\Omega_m^2 + \Omega_p^2\cos{(2\Omega t + \phi_t)} \right]\delta x - \Gamma_m \delta\dot{x} + \frac{\hbar G \bar{\alpha}}{m}\left(\delta\alpha + \delta\alpha^*\right) \\
\delta\dot{\alpha} & = & \left[i\bar{\Delta} - \frac{\kappa}{2}\right]\delta\alpha + iG\bar{\alpha}\delta x + \sqrt{\frac{\kappa_e}{2}}S_p.
\end{eqnarray}
where $\Omega_p^2 = k_p/m$ and $\phi_t$ considers an additional possible phase shift.
We solve these equations again with the Ansatz
\begin{eqnarray}
\delta\alpha & = & a_-e^{-i\Omega t} + a_+e^{+i\Omega t}\\
\delta\alpha^* & = & a_-^*e^{+i\Omega t} + a_+^* e^{-i\Omega t}\\
\delta x & = & x_1e^{-i\Omega t} + x_1^* e^{+i\Omega t}
\end{eqnarray}
and the identity
\begin{equation}
\cos{(2\Omega t + \phi_t)} = \frac{1}{2}\left[e^{+i2\Omega t}e^{+i\phi_t} + e^{-i2\Omega t}e^{-i\phi_t}\right].
\end{equation}
Using rotating wave approximation and algebra yields the solution
\begin{eqnarray}
x_1 & = & \hbar G \bar{\alpha} \chi_m^\mathrm{eff}\left[\frac{\chi_c - \left(\chi_m^\mathrm{eff}\right)^*\chi_c^*\frac{m\Omega_p^2}{2}e^{-i\phi_t}}{1-\left|\chi_m^\mathrm{eff}\right|^2\frac{m^2\Omega_p^4}{4}}\right]\sqrt{\frac{\kappa_e}{2}}S_0\\
a_- & = & \chi_c\left[1 + i2m\Omega_m g^2 \chi_m^\mathrm{eff}\frac{\chi_c - \left(\chi_m^\mathrm{eff}\right)^*\chi_c^*\frac{m\Omega_p^2}{2}e^{-i\phi_t}}{1-\left|\chi_m^\mathrm{eff}\right|^2\frac{m^2\Omega_p^4}{4}}\right]\sqrt{\frac{\kappa_e}{2}}S_0.
\end{eqnarray}
This can be significantly simplified for drives on one of the sidebands, i.e., for $\omega_d = \omega_c \pm \Omega_m$ and a probe very close to the cavity resonance $\omega_p \sim \omega_c$.

\subsubsection{Drive on the red sideband}

\begin{figure}[h]
	\centering {\includegraphics[trim={0cm 10cm 0cm 1cm},clip=True,scale=0.55]{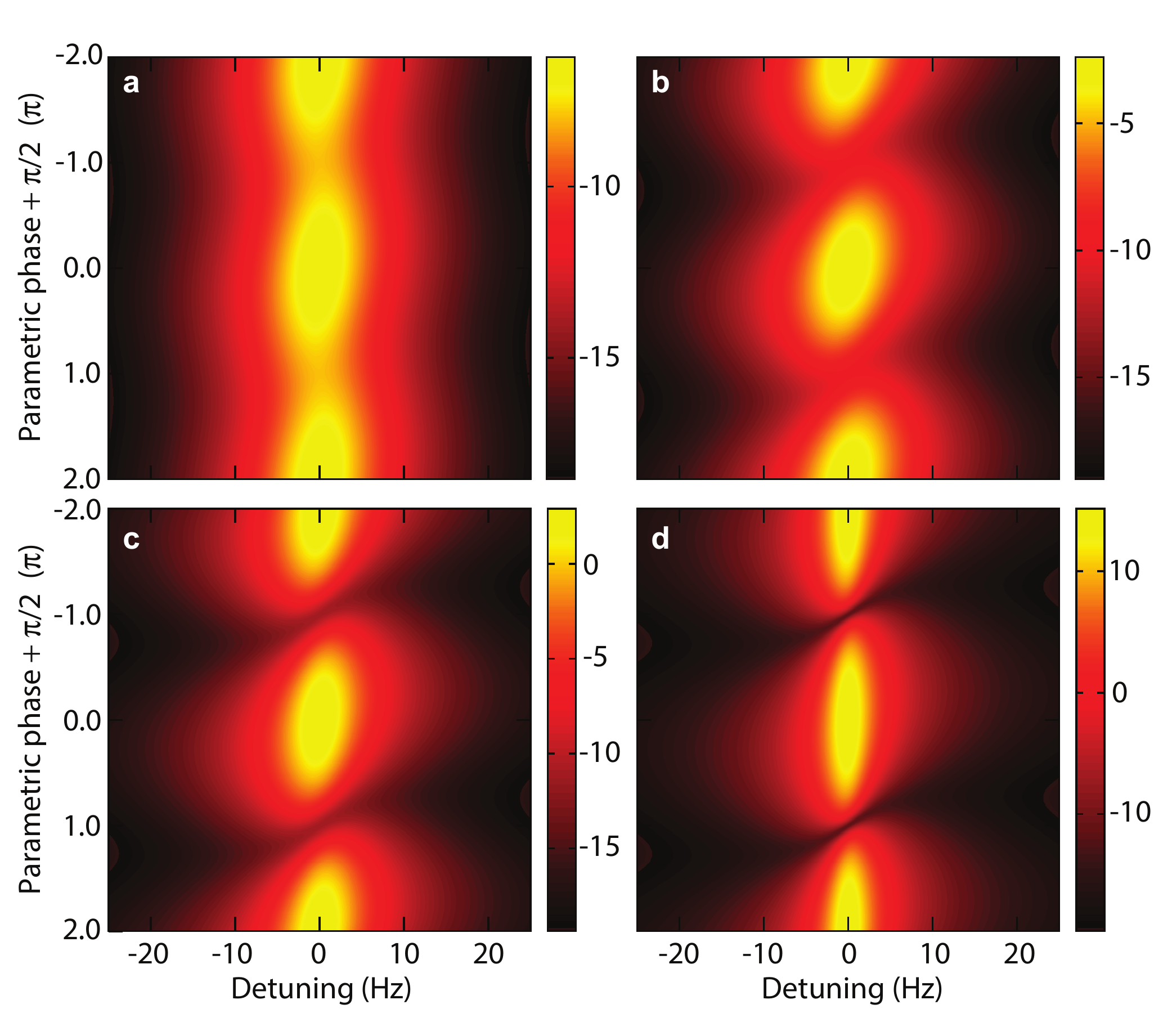}}
	\caption{\textsf{\textbf{Calculated OMIT microwave transmission with mechanical parametric driving.} Plots show the microwave transmission signal $S_{21}$ in dB vs parametric offset phase $\phi_t - \pi/2$ and vs detuning from the mechanical resonance frequency $\Omega_m$ for four different parametric modulation amplitudes. \textbf{a} $V_{2\Omega}/V_t = 0.12$, \textbf{b} $V_{2\Omega}/V_t = 0.46$, \textbf{c} $V_{2\Omega}/V_t = 0.72$ and \textbf{d} $V_{2\Omega}/V_t = 0.93$. The transmission shows a $2\pi$-periodicity in parametric modulation phase. The calculation parameters were chosen close to the experimental device with $\kappa_i = 2\pi\cdot 370\,$kHz, $\kappa_e = 2\pi\cdot5.6\,$MHz, $\Omega_m = 2\pi\cdot 1.4315\,$MHz, $C = 0.5$ and $Q_m = 146000$, the numbers at the color bars represent the transmission parameter $S_{21}$ in dB. In \textbf{c} and \textbf{d} a microwave signal going to the device experiences a net amplification.}}
	\label{fig:OMITGainTheory}
\end{figure}
\begin{figure}[h]
	\centering {\includegraphics[trim={0cm 20cm 0cm 1cm},clip=True,scale=0.75]{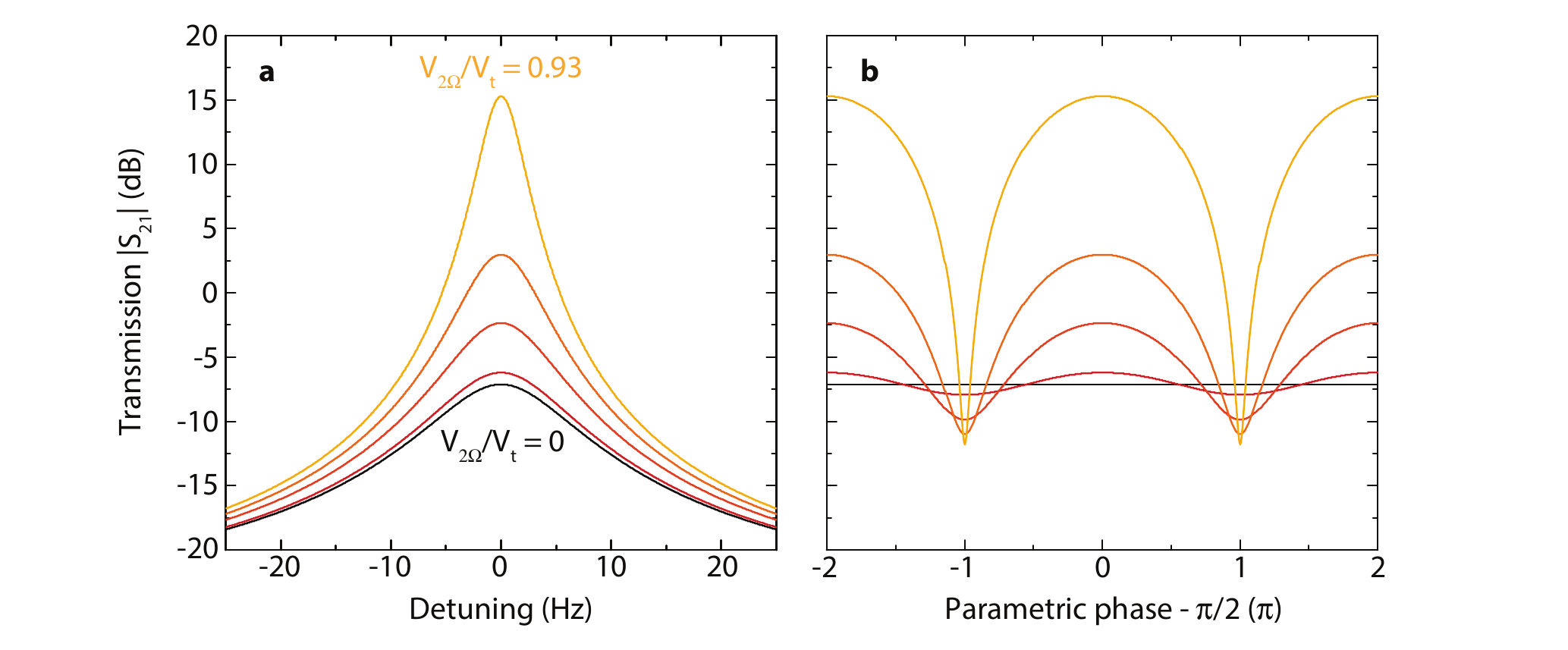}}
	\caption{\textsf{\textbf{Maximum gain and phase sensitivity of the amplification.} In \textbf{a} we show the maximum gain extracted from the calculated data in Fig.~\ref{fig:OMITGainTheory}, the lowest curve shows the OMIT signal without parametric drive. Note that for each detuning, the absolute phase to reach the shown maximum is different. In \textbf{b} we plot the phase dependence of the transmitted signal for zero detuning, which corresponds to vertical linecuts through the panels of Fig.~\ref{fig:OMITGainTheory}.}}
	\label{fig:OMITmax}
\end{figure}

When the cavity drive is set to the red sideband $\omega_d = \omega_c - \Omega_m$ and the probe tone is sweeping only very close to the cavity resonance frequency $\omega_p =\omega_c + \Delta_m$ with $\Delta_m \ll \kappa$, we can significantly simplify the equations.
The cavity susceptibility becomes $\chi_c = \chi_c^* = 2/\kappa$ and the effective mechanical susceptibility becomes
\begin{equation}
\chi_m^\mathrm{eff} = -\frac{1}{m\Omega_m}\frac{1}{2\Delta_m + i\Gamma_\mathrm{eff}}.
\end{equation}
After introducing $\chi_m^\mathrm{eff} = |\chi_m^\mathrm{eff}|e^{i\varphi_m}$ and the parameter
\begin{equation}
B = \frac{\Omega_p^2}{2\Omega_m}\frac{1}{\sqrt{\Gamma_\mathrm{eff}^2 + 4\Delta_m^2}} = |\chi_m^\mathrm{eff}|\frac{m\Omega_p^2}{2}
\end{equation}
we can rewrite the intracavity amplitude as
\begin{equation}
a_- = \frac{2}{\kappa}\left[1 - i\frac{C\Gamma_m}{2\Delta_m + i\Gamma_\mathrm{eff}}\frac{1 - Be^{-i(\phi_t + \varphi_m)}}{1 - B^2}\right]\sqrt{\frac{\kappa_e}{2}}S_0
\end{equation}
and the transmission as
\begin{equation}
S_{21} = \frac{\kappa_i}{\kappa} + i\frac{\kappa_e}{\kappa}\frac{C\Gamma_m}{2\Delta_m + i\Gamma_\mathrm{eff}}\frac{1-Be^{-i(\phi_t + \varphi_m)}}{1 - B^2}.
\label{eqn:S21OMITGain}
\end{equation}
For the minimum and maximum transmitted power exactly on resonance we get
\begin{equation}
|S_{21}|^2 = \frac{\kappa_i^2}{\kappa^2} + \frac{C_p\Gamma_m}{\Gamma_\mathrm{eff}^2}\left[2\frac{\kappa_i \kappa_e}{\kappa^2}\Gamma_\mathrm{eff} + \frac{\kappa_e^2}{\kappa^2}C_p\Gamma_m\right],
\end{equation}
which is the same equation as without parametric drive, but with a parametrically enhanced/reduced cooperativity
\begin{equation}
C_p = \frac{C}{1\pm B_0}
\end{equation} 
with $B_0 = \Omega_p^2/2\Omega_m\Gamma_m$
The net microwave power gain in this regime is given by $G_\mathrm{mw} = |S_{21}|_\mathrm{max}^2 - 1$.
It is interesting to notice that the parameter $B$ we introduced here, corresponds exactly to the voltage ratio $V_{2\Omega}/V_t'(\Gamma_\mathrm{eff})$, cf. Sec.~\ref{sec:TPMA}, but with a threshold voltage determined by the effective mechanical linewidth.
This means that the parametric instability regime onset is modified by the optomechanical interaction.
In Fig.~\ref{fig:OMITGainTheory} we plot the result of Eq.~(\ref{eqn:S21OMITGain}) for four different parametric modulation strengths.
The calculation parameters are chosen to be close to the device parameters, i.e., $\kappa_i = 2\pi\cdot 370\,$kHz, $\kappa_e = 2\pi\cdot5.6\,$MHz, $\Omega_m = 2\pi\cdot 1.4315\,$MHz, $C = 0.5$ and $Q_m = 146000$, where the latter is adjusted to the value we expect for the corresponding sideband drive power, cf. Fig.~\ref{fig:Geff}\textbf{a}.
For small modulation $V_{2\Omega}/V_t = 0.12$ as shown in \textbf{a}, the OMIT signal is only slightly distorted from the signal without parametric modulation, but the phase sensitivity of the amplifier becomes already apparent.
For larger parametric modulations, the maximum gain increases until it reaches $\sim 15\,$dB for $V_{2\Omega}/V_t = 0.93$ as shown in \textbf{d}.
Linecuts for the phase-dependence at zero detuning and the lines of maximum gain for each power are shown in Fig.~\ref{fig:OMITmax}\textbf{a} and \textbf{b}.
Note that the phase periodicity here is $2\pi$, which is a consequence of including the parametric phase into the parametric drive in the theoretical treatment here instead of the driving force.
In addition, we have to consider a formal phase lag of $\pi/2$ induced by the phase of the cavity response function, which on resonance vanishes, while the phase of a directly driven mechanical oscillator on resonance is $-\pi/2$.

\subsubsection{Drive on the blue sideband}

\begin{figure}[h]
	\centering {\includegraphics[trim={0cm 20cm 0cm 1cm},clip=True,scale=0.75]{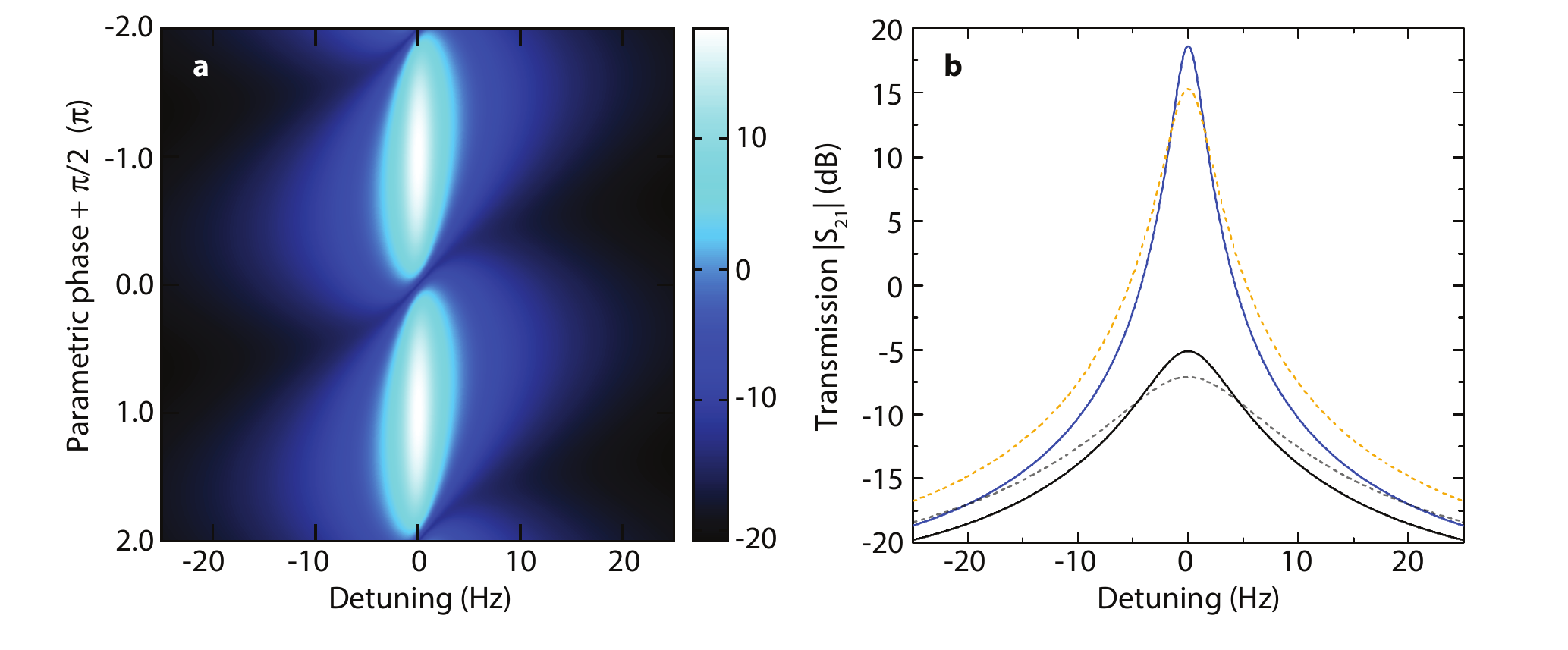}}
	\caption{\textsf{\textbf{Phase-sensitive amplification with a drive on the blue sideband.} In \textbf{a} we show the cavity transmission $|S_{21}|$ in dB for a drive on the blue sideband and a parametric excitation amplitude $V_{2\Omega}/V_t \approx 0.93$. All other parameters are as given in the caption of Fig.~\ref{fig:OMITGainTheory}. In \textbf{b} we show as solid lines the window of maximum transmission without parametric drive (black) and with parametric drive (blue) as extracted from \textbf{a}. The dashed lines show the equivalent curves for a red detuned drive.}}
	\label{fig:AmpOMIA}
\end{figure}

On the blue sideband, we get $\chi_m^\mathrm{eff} = -|\chi_m^\mathrm{eff}|e^{i\varphi_m}$ and thus
\begin{equation}
a_- = \frac{2}{\kappa}\left[1 + i\frac{C\Gamma_m}{2\Delta_m + i\Gamma_\mathrm{eff}'}\frac{1 + Be^{-i(\phi_t + \varphi_m)}}{1 - B^2}\right]\sqrt{\frac{\kappa_e}{2}}S_0
\end{equation}
for the intracavity field.
The transmission parameter becomes
\begin{equation}
S_{21} = \frac{\kappa_i}{\kappa} - i\frac{\kappa_e}{\kappa}\frac{C\Gamma_m}{2\Delta_m + i\Gamma_\mathrm{eff}'}\frac{1 + Be^{-i(\phi_t + \varphi_m)}}{1 - B^2}
\end{equation}
One example for the gain obtained when driving on the blue sideband is given in Fig.~\ref{fig:AmpOMIA}\textbf{a}.
The phase sensitivity and the total gain is comparable to the values obtained for a drive on the red sideband for a similar value of $V_{2\Omega}/V_t = 0.93$ with an additional phase shift of $pi$ in the parametric phase dependence.
In \textbf{b}, the maximum microwave transmission is shown as blue line and the bare transmission signal without parametric drive as black line for this parameter regime.
For comparison, the equivalent data for the red sideband drive are shown as dashed lines.
We note that although the relative parametric pump strength is comparable for the red and the blue detuned drive here, the absolute numbers are different, due to the smaller effective mechanical linewidth in the blue-detuned case.

\section{Parametric microwave amplification - Measurement routine and data processing}

Both, the measurement routine and the data processing are done in full analogy with the mechanical amplitude amplification.
Instead of sweeping the phase, we detune the parametric drive tone by $\sim0.1\,$Hz from the frequency difference between the sideband drive and the probe tone .
Then, we track the transmission of a probe tone vs time with a network analyzer.
The resulting oscillatory transmission curves of the amplitude are fitted with a function as given in Eq.~{\ref{eqn:GainFit}}, from which we extract maximum and minimum transmission.
To normalize the signal, we calculate the nominal complex background value at the corresponding frequency from the cavity fit and normalize the transmission with it.

\end{document}